\documentclass[aps,prd,superscriptaddress,notitlepage,showpacs,10]{revtex4-1}

\usepackage{graphicx}
\usepackage{amsmath}
\usepackage{amssymb}
\usepackage{color}
\usepackage{bm}

\begin{document}

\title{Radiative corrections to chiral separation effect in QED}

\date{July 4, 2013}

\preprint{UWO-TH-13/6}

\author{E. V. Gorbar}
\affiliation{Department of Physics, Taras Shevchenko National Kiev University, Kiev, 03022, Ukraine}
\affiliation{Bogolyubov Institute for Theoretical Physics, Kiev, 03680, Ukraine}

\author{V. A. Miransky}
\affiliation{Department of Applied Mathematics, Western University, London, Ontario N6A 5B7, Canada}

\author{I. A. Shovkovy}
\affiliation{School of Letters and Sciences, Arizona State University, Mesa, Arizona 85212, USA}

\author{Xinyang Wang}
\affiliation{Department of Physics, Arizona State University, Tempe, Arizona 85287, USA}

\begin{abstract}
We calculate the leading radiative corrections to the axial current in the chiral separation
effect in dense QED in a magnetic field. Contrary to the conventional wisdom suggesting 
that the axial current should be exactly fixed by the chiral anomaly relation and is described 
by the topological contribution on the lowest Landau level in the free theory, we find in fact 
that the axial current receives nontrivial radiative corrections. The direct calculations performed 
to the linear order in the external magnetic field show that the nontrivial radiative corrections to 
the axial current are provided by the Fermi surface singularity in the fermion propagator 
at nonzero fermion density. 
\end{abstract}


\maketitle

\section{Introduction}

Recently there was significant interest in the dynamics of relativistic matter in a magnetic
field. Assuming that QCD topological fluctuations produce local ${\cal P}$ and ${\cal CP}$-odd 
states \cite{Kharzeev:2004ey} leading to a chiral chemical potential $\mu_5$, it was suggested 
that there exists a nondissipative electric current $\mathbf{j}=e^2\mathbf{B}\mu_5/(2\pi^2)$ 
in relativistic matter in a magnetic field $\mathbf{B}$ \cite{Kharzeev:2007tn,Kharzeev:2007jp,
Fukushima:2008xe}. This phenomenon is known in the literature as the chiral magnetic effect 
(CME). (For a recent review see Ref.~\cite{Fukushima:2012vr}.) Moreover, the charge-dependent 
correlations and flow, observed in heavy-ions collisions at the RHIC \cite{collisions,Adamczyk:2013kcb,
Wang:2012qs,Ke:2012qb} and LHC \cite{Selyuzhenkov:2011xq}, appear to be in a qualitative 
agreement with the predictions of the CME \cite{Voloshin:2004vk,Kharzeev:2009fn}.

Unlike the chiral chemical potential, which is a rather exotic quantity and not so well defined theoretically, 
the chemical potential $\mu$ (associated, for example, with conserved electric or baryon charges) is 
common in many physical systems. It was shown in Refs.~\cite{Vilenkin:1980ft,Zhitnitsky,Newman} that 
a nondissipative axial current $\mathbf{j}_5=e\mathbf{B}\mu/(2\pi^2)$ exists in the equilibrium 
state of noninteracting massless fermion matter in a magnetic field. This effect is known as the 
chiral separation effect (CSE) in the literature. (For a brief review, see Sec. 2 in Ref.~\cite{Fukushima:2012vr}.) 
In fact, as suggested in Refs.~\cite{chiral-shift-3,Burnier:2011bf}, the CSE may 
lead to a chiral charge separation (i.e., effectively inducing a nonzero chiral chemical 
potential $\mu_5$) and, thus, trigger the CME even in the absence of topological fluctuations 
in the initial state.

The approach in Refs.~\cite{Zhitnitsky,Newman} was based on the use of the operator 
form of the chiral anomaly relation \cite{ABJ}. It is well known that the corresponding relation calculated 
at one-loop order is exact and, as such, it cannot get any higher-order radiative corrections \cite{Adler:1969er}. 
Therefore, it was argued in \cite{Zhitnitsky,Newman} that like the chiral anomaly, the one-loop result for the 
axial current density $\mathbf{j}_5=e\mathbf{B}\mu/(2\pi^2)$ should be exact as well.

Since the fermion propagator in a magnetic field depends nonlinearly 
on the magnetic field, the linear dependence of the axial current on $\mathbf{B}$ calls 
for a physical explanation. Using an expansion over the Landau levels, it was shown 
in Ref.~\cite{Zhitnitsky} that the axial current $\mathbf{j}_5=e\mathbf{B}\mu/(2\pi^2)$ 
is topological in nature (see also Ref.~\cite{Basar-Dunne} for a nice exposition and
some details) and is defined by the fermion number density on the lowest Landau level (LLL). 
Moreover, it was shown \cite{Zhitnitsky} that a similar result holds even 
for massive fermions at finite temperature $T$, where the axial current equals 
$\mathbf{j}_5=e\mathbf{B}n_L(m,T)/(2\pi)$ and $n_L(m,T)$ is the effective one-dimensional 
(along the direction of magnetic field) fermion number density on the LLL. At zero temperature 
the axial current is given by $\mathbf{j}_5=e\mathbf{B}\sqrt{\mu^2-m^2}/(2\pi^2)$. Of course, 
in the chiral limit $m \to 0$ this reduces to the same expression for the axial current as derived from 
the chiral anomaly. Note, however, that the connection between the induced axial current 
and the anomaly relation is not obvious beyond the chiral limit.

The chiral anomaly is exact as an operator relation, but it contains the divergence of the axial current rather 
than the current itself. Consequently, to get the axial current from the chiral anomaly one should ``integrate" 
the anomaly and calculate the ground state expectation value of the corresponding operator. Then, the 
question concerning an ``integration constant" in the induced axial current and its dependence on interactions
naturally arises. Until now, no conclusive answer to this question was given (e.g., see the discussion in 
Ref.~\cite{Fukushima:2012vr}). 

The first studies of the interactions effects were done in 
Refs.~\cite{chiral-shift-1,chiral-shift-2,chiral-shift-3,Fukushima:2010zza} in the framework of the 
dense Nambu--Jona-Lasinio (NJL) model in a magnetic field. Using the Schwinger--Dyson 
equation for the fermion propagator, it was found \cite{chiral-shift-1,chiral-shift-2,chiral-shift-3} 
that the four-fermion interactions generate a chiral shift parameter $\Delta$. In the chiral limit, 
this parameter determines a relative shift of the momenta in the dispersion relations for opposite chirality 
fermions $k^3 \to k^3 \pm\Delta$, where the momentum $k^3$ is directed along the magnetic field. The presence of the
chiral shift parameter leads to an additional dynamical contribution in the axial current. Unlike the topological
contribution in the axial current at the LLL, the dynamical one affects the fermions in all Landau levels, including those
around the whole Fermi surface. Further, it was explicitly checked in Ref.~\cite{chiral-shift-2} that although the axial current gets 
corrections due to the NJL interactions, the chiral anomaly does not. 

Since the NJL model is nonrenormalizable and the chiral anomaly is intimately connected with ultraviolet 
divergencies, in order to reach a solid conclusion about the presence or absence of
higher-order radiative corrections to the axial current, one should consider them   
in a renormalizable model. In the present paper, assuming that the magnetic field $\mathbf{B}$ is weak 
and using the expansion in powers of $\mathbf{B}$ up to linear order, the leading radiative 
corrections to the axial current in QED are calculated. We find that they do not vanish and attribute 
this result to the singularities in the fermion propagator at the Fermi surface. On the technical side, the
$i\epsilon\,\mbox{sign}(k_0)$ prescription in the fermion propagator, which is the only thing that distinguishes a
chemical potential from the time component $A_0$ of the photon field, plays a crucial role in deriving
this result.

This paper is organized as follows. In Sec.~\ref{Sec:Model} we introduce the model and set up the 
notation. Also, we discuss some properties of the fermion propagator and the one-loop self-energy in the presence 
of an external magnetic field and a nonzero density. The calculation of the leading radiative corrections
to the axial current is presented in Sec.~\ref{Sec:axial-current}. We start from the formal definition of the current
in terms of the fermion propagator, use its systematic expansion in powers of the magnetic field, and finally 
perform the explicit calculations. Our discussion of the results and conclusions are given in Sec.~\ref{Sec:Conclusion}.
A new form of the Schwinger parametrization for the fermion propagator in the case of a nonzero magnetic field 
and a nonzero chemical potential, utilized in the main part of the paper, is presented in 
Appendix~\ref{App:proper-time-rep}. The details of the calculations of the radiative corrections to the axial 
current are given in Appendix \ref{App:calculation-j35}.

\section{Fermion self-energy in a magnetic field}
\label{Sec:Model}

The Lagrangian density of QED in a magnetic field is given by
\begin{equation}
{\cal L} = -\frac{1}{4}F^{\mu\nu}F_{\mu\nu}+\bar{\psi}\left( i\gamma^{\nu}{\cal D}_{\nu}+\mu
\gamma^0-m\right)\psi +\delta_2\bar{\psi}( i \gamma^{\nu} \partial_{\nu}+\mu\gamma^{0}+e A^{\rm ext}_{\nu} \gamma^{\nu}
)\psi-\delta_m\bar{\psi}\psi,
\label{Lagrangian}
\end{equation}
where $\mu$ is the fermion chemical potential, the last two terms are counterterms (we use the notation of 
Ref.~\cite{Peskin}, but with the opposite sign of the electric charge, $e\to-e$), and the covariant derivative is 
${\cal D}_{\mu} =\partial_{\mu}-i e A_{\mu}-i e A^{\rm ext}_{\mu}$. Without the loss of generality, we assume that the
external magnetic field $\mathbf{B}$ points in the $+x^3$ direction and is described by the vector potential in the Landau gauge,
$A_\mu^{\rm ext}=\left(0,0,Bx_1,0\right)$. Note that the counterterms include the chemical potential $\mu$ and the external field
$A^{\rm ext}_{\mu}$.

To leading order in the coupling constant $\alpha=e^2/(4\pi)$, the fermion self-energy in QED is 
given by 
\begin{equation}
\Sigma(x,y)=-4i\pi\alpha\gamma^\mu S(x,y) \gamma^\nu D_{\mu\nu}(x-y),
\label{self-energy}
\end{equation}
where $S(x,y)$ is the free fermion propagator in magnetic field and $D_{\mu\nu}(x-y)$ is the free 
photon propagator. 

As is well known, the fermion propagator $S(x,y)$ in the presence of an external magnetic field is
not translation invariant. It can be written, however, in a form of an overall Schwinger phase 
(breaking the translation invariance) and a translation invariant function \cite{Schwinger:1951nm}, 
i.e.,
\begin{equation}
S(x,y) = \exp\left[i\Phi(x,y)\right]\bar{S}(x-y),
\label{S-Schwinger-phase}
\end{equation}
where the Schwinger phase equals $\Phi(x,y)=-eB(x_1+y_1)(x_2-y_2)/2$ in the Landau gauge. 
The Fourier transform of $\bar{S}(x-y)$ is presented in Eq.~(\ref{Fourier-tranlation-inv-S}) in 
Appendix~\ref{App:proper-time-rep}. The expression in Eq.~(\ref{self-energy}) implies that 
the self-energy $\Sigma(x,y)$ has an analogous representation
\begin{equation}
\Sigma(x,y) = \exp\left[i\Phi(x,y)\right]\bar{\Sigma}(x-y) ,
\label{Sigma-Schwinger-phase}
\end{equation}
with the same Schwinger phase as in the propagator.

In this study we use the photon propagator in the Feynman gauge. In momentum space, 
it reads
\begin{equation}
D_{\mu\nu}(q)=-i\frac{g_{\mu\nu}}{q^2_{\Lambda}} \equiv  
-i\left(\frac{g_{\mu\nu}}{q_0^2-\mathbf{q}^2-m_\gamma^2+i\epsilon}
-\frac{g_{\mu\nu}}{q_0^2-\mathbf{q}^2-\Lambda^2+i\epsilon} \right).
\label{photon-propagator}
\end{equation}
Here we introduced a nonzero photon mass $m_{\gamma}$ which serves as an infrared regulator 
at the intermediate stages of calculations. Of course, none of the physical observables should 
depend on this parameter (see Sec.~\ref{Sec:Conclusion} below).
(Note that since the classical paper of Stueckelberg \cite{Stueckelberg:1957zz}, 
it is well known that, unlike non-Abelian theories, introducing a photon mass causes no problems
in an Abelian gauge theory, such as QED.) We will see in Sec.~\ref{Sec:axial-current} that the leading radiative 
corrections are logarithmically divergent in the ultraviolet region. As in Ref.~\cite{Adler:1969er},
we find that the Feynman regularization of the photon propagator (\ref{photon-propagator})
with ultraviolet regularization parameter $\Lambda$ presents the most convenient way of
regularizing the theory.

The Fourier transform of the translation invariant function $\bar{\Sigma}(x-y)$ is given 
by the following expression:
\begin{equation}
\bar{\Sigma} (p) = -4i\pi\alpha \int\frac{ d^4 k}{(2\pi)^4}
\gamma^{\mu}\,\bar{S} (k)\gamma^{\nu} D_{\mu\nu}(k-p), 
\label{self-energy-momentum}
\end{equation}
where $\bar{S}(k)$ is the Fourier transform of the translation invariant part of the 
fermion propagator and $D_{\mu\nu}(q)$ is the photon propagator (\ref{photon-propagator}). 

To linear order in $B$, the translation invariant part of the free fermion propagator in the 
momentum representation has the following structure:
\begin{equation}
\bar{S}(k) = \bar{S}^{(0)}(k) +\bar{S}^{(1)}(k) +\cdots,
\end{equation}
where $\bar{S}^{(0)}$ is the free fermion propagator in the absence of magnetic field
and $\bar{S}^{(1)}$ is the linear in the magnetic field part. Both of them are derived 
in Appendix~\ref{App:proper-time-rep} by making use of a generalized Schwinger 
parametrization when the chemical potential is nonzero. The final expressions for
$\bar{S}^{(0)}$  and $\bar{S}^{(1)}$ can be also rendered in the following equivalent form:
\begin{equation}
\bar{S}^{(0)}(k) = i \frac{(k_0+\mu)\gamma^0- \mathbf{k}\cdot\bm{\gamma}+m}
{(k_0+\mu+i\epsilon\, \mbox{sign}(k_0))^2-\mathbf{k}^2-m^2}
\label{free-term}
\end{equation}
and 
\begin{equation}
\bar{S}^{(1)}(k) =  - \gamma^1\gamma^2 eB \frac{(k_0+\mu)\gamma^{0}-k_3 \gamma^3+m}
{\left[ (k_0+\mu+i\epsilon\, \mbox{sign}(k_0))^2-\mathbf{k}^2 -m^2\right]^2 }.
\label{linear-term}
\end{equation}
The self-energy at zero magnetic field
\begin{equation}
\bar{\Sigma}^{(0)}(p)=-4i\pi \alpha \int\frac{d^4k}{(2\pi)^4}\gamma^\mu \bar{S}^{(0)}(k) \gamma^\nu D_{\mu\nu}(p-k)
\label{self-energy-momentum-space-B0}
\end{equation}
determines the counterterms $\delta_2$ and $\delta_m$ in Eq.~(\ref{Lagrangian}). 
To calculate the self-energy (\ref{self-energy-momentum-space-B0}),
we will use the generalized Schwinger parametrization of the fermion propagator $\bar{S}^{(0)}(k)$,
see Eq.~(\ref{S0-vac-plus-matter}) in Appendix~\ref{App:proper-time-rep}. Such a representation allows a natural  
separation of the propagator (as well as the resulting self-energy) into the ``vacuum" and ``matter" parts. The 
former is very similar to the usual vacuum self-energy in QED in the one-loop approximation. The only difference 
will be the appearance of $p_0+\mu$ instead of $p_0$. The matter part is an additional contribution that comes
from the $\delta$-function contribution in Eq.~(\ref{S-prop-time-eB0}). Unlike the vacuum part, the matter one has no 
ultraviolet divergences and vanishes when $|\mu|<m$. 

The explicit expression for the vacuum part reads
\begin{equation}
\bar{\Sigma}^{(0)}_{\rm vac} (p) =\frac{\alpha}{2\pi}
\int_0^1 dx \left\{2m-x\left[(p_0+\mu)\gamma^{0}-\mathbf{p}\cdot\bm{\gamma}\right]\right\}
\ln\frac{x \Lambda^2}{(1-x) m^2 + x m_\gamma^2 - x(1-x)\left[(p_0+\mu)^2-\mathbf{p}^2\right]}.
\label{self-energy-vacuum}
\end{equation}
Note that, while the integral over $x$ can be easily calculated, we keep the result in this more 
compact form. We see that the self-energy (\ref{self-energy-vacuum}) becomes identical with 
the well-known vacuum self-energy in QED in the Feynman gauge after performing the 
substitution $p_0+\mu\to p_0$ \cite{Peskin}. Further, using Eq.~(\ref{self-energy-vacuum}), 
we find that the counterterms in (\ref{Lagrangian}) are defined as follows \cite{Peskin}:
\begin{eqnarray}
\delta_2 &=& \frac{d\bar{\Sigma}^{(0)}_{\rm vac}(p)}{d\!\! \not{\!P}}\Big|_{\not{P}=m}= -\frac{\alpha}{2\pi}\left(\frac{1}{2}\ln\frac{\Lambda^2}{m^2}+\ln\frac{m_\gamma^2}{m^2}+\frac{9}{4}\right) ,
\label{wave-function-renormalization-constant} \\
\delta m &=& m-m_0 = \bar{\Sigma}^{(0)}_{\rm vac}(p)\Big|_{\not{P}=m} =\frac{3\alpha}{4\pi} m \left(\ln\frac{\Lambda^2}{m^2}+\frac{1}{2}\right),
\label{mass-renormalization-constant}
\end{eqnarray}
where $P=(p_0+\mu,\mathbf{p})$. Note that the fermion wave function renormalization constant is defined 
as follows: $Z_2=1+\delta_2$.

For completeness, let us calculate the additional matter part of the self-energy due to the filled fermion states given by
\begin{equation}
\bar{\Sigma}^{(0)}_{\rm mat} (p) = -\frac{i\alpha}{\pi^2} \int_{-\mu}^{0} d k_0 \int  d^3\mathbf{k}\, 
\frac{(k_0+\mu)\gamma^{0}-\mathbf{k}\cdot\bm{\gamma}-2m}{(k_0-p_0)^2-(\mathbf{k}-\mathbf{p})^2}\,\,
\delta \left[(k_0+\mu)^2-\mathbf{k}^2 -m^2 \right].
\end{equation}
After performing the integration over the energy and spatial angular coordinates, we find
\begin{eqnarray}
\bar{\Sigma}^{(0)}_{\rm mat} (p) &=&
-\frac{\alpha}{\pi}\int_{0}^{\sqrt{\mu^2-m^2}}\frac{k dk}{|\mathbf{p}|}\Bigg\{\frac{1}{2}\left(\gamma^0-\frac{2m}{\sqrt{k^2+m^2}}\right) 
\ln\frac{(p_0+\mu-\sqrt{m^2+k^2})^2-(k-|\mathbf{p}|)^2}{(p_0+\mu-\sqrt{m^2+k^2})^2-(k+|\mathbf{p}|)^2}
\nonumber\\
&-&\frac{k (\mathbf{p}\cdot\bm{\gamma})}{|\mathbf{p}|\sqrt{m^2+k^2}} 
\left(1+\frac{k^2+\mathbf{p}^2-(p_0+\mu-\sqrt{m^2+k^2})^2}{4k|\mathbf{p}|} 
\ln\frac{(p_0+\mu-\sqrt{m^2+k^2})^2-(k-|\mathbf{p}|)^2}{(p_0+\mu-\sqrt{m^2+k^2})^2-(k+|\mathbf{p}|)^2}\right)
\Bigg\}.
\end{eqnarray}
While the remaining integral over the absolute value of the momentum $k$ can be also performed, 
the result will take a rather complicated form that will not add any clarity.

The linear in the magnetic field correction to the translation invariant part of the fermion self-energy in a magnetic field reads
\begin{equation}
\bar{\Sigma}^{(1)}(p)=-4i\pi \alpha \int\frac{d^4k}{(2\pi)^4}\gamma^\mu \bar{S}^{(1)}(k) \gamma^\nu D_{\mu\nu}(p-k).
\label{self-energy-momentum-space-B1}
\end{equation}
This correction, which in particular contains a chiral shift parameter term, has been recently analyzed 
in Ref. \cite{Gorbar:2013uga}. We use this expression for $\bar{\Sigma}^{(1)}(p)$ in the derivation 
of the leading corrections in the axial current in Sec.~\ref{Sec:axial-current} below.

\section{The leading radiative corrections to the axial current}
\label{Sec:axial-current}

The renormalization group invariant axial current density, which is a quantity of 
the principal interest in the present paper, is given by
\begin{equation}
\langle j_{5}^{3}\rangle = -Z_2\mbox{tr}\left[\gamma^3\gamma^5 G(x,x)\right],
\label{j53-general}
\end{equation}
where $G(x,y)$ is the full fermion propagator and $Z_2 = 1 + \delta_2$ is the wave function 
renormalization constant of the fermion propagator, cf. Eq.~(\ref{Lagrangian}).

To the first order in the coupling constant $\alpha=e^2/(4\pi)$, the propagator reads
\begin{equation}
G(x,y) = S(x,y)+i\int d^4 u d^4 v S(x,u) \Sigma(u,v) S(v,y)  + i\int d^4 u d^4 v S(x,u) \Sigma_{\rm ct}(u,v) S(v,y),
\label{gxy-expansion}
\end{equation}
where $S(x,y)$ is the free fermion propagator in the magnetic field, $\Sigma(u,v)$ is 
the one-loop fermion self-energy, and $\Sigma_{\rm ct}(u,v)$ is the counterterm
contribution to the self-energy. The structure of the counterterm contribution is 
determined by the last two terms in the Lagrangian density (\ref{Lagrangian}). 

\begin{figure}[t!]
\includegraphics[width=0.45\textwidth]{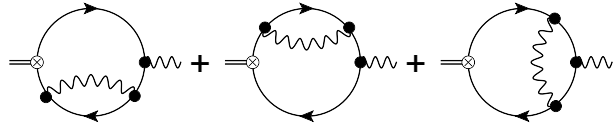}
\caption{The leading radiative corrections to the axial current in the linear 
in magnetic field approximation. Solid and wavy lines correspond to the fermion and photon 
propagators, respectively. Double solid lines describe the axial current insertions and the external 
wavy lines attached to the fermion loops indicate the insertions of the external gauge field.}
\label{fig-correlator}
\end{figure}

In this paper, we make use of the weak magnetic field expansion in the calculation 
of the axial current density. Such an expansion is straightforward to obtain from the 
general expression in Eq.~(\ref{j53-general}) and the representation (\ref{gxy-expansion}) 
for the fermion propagator. For the fermion propagator to linear in $B$ order, we have
\begin{equation}
S(x,y) = \bar{S}^{(0)}(x-y) + i e\int d^4z\,\bar{S}^{(0)}(x-z)\gamma^{\nu}\bar{S}^{(0)}(z-y)A^{\rm ext}_{\nu}(z).
\label{free-propagator-magnetic-1}
\end{equation}
Further, by making use of Eq.~(\ref{free-propagator-magnetic-1}), the weak field expansion of the 
self-energy follows from the definition in Eq.~(\ref{self-energy}). (Note that the photon 
propagator is independent of the magnetic field to this order.) Combining all pieces 
together, we can find the complete expression for the leading radiative corrections 
to the axial current (\ref{j53-general}) in the approximation linear in the magnetic field. 
In this framework, the diagrammatical representation for the leading radiative corrections 
to the axial current is shown in Fig.~\ref{fig-correlator} (for simplicity, we do not display 
the contributions due to counterterms) \cite{footnote-neutrality}.
   
Instead of using the expansion for the free propagator in Eq.~(\ref{free-propagator-magnetic-1}), we find it 
much more convenient to utilize the Schwinger form of the fermion propagator (\ref{S-Schwinger-phase}), 
which consists of a simple phase, that breaks the translation invariance, and a translation invariant function. 
Taking into account that the Schwinger phase $\Phi(x,y)$ is linear in magnetic field, we arrive at the following 
alternative form of the weak field expansion of the fermion propagator in the linear in $B$ approximation: 
\begin{equation}
S(x,y) = \bar{S}^{(0)}(x-y)+i\Phi(x,y)\bar{S}^{(0)}(x-y)+\bar{S}^{(1)}(x-y),
\label{free-propagator-magnetic-2}
\end{equation}
where $\bar{S}^{(0)}(x-y)$ and $\bar{S}^{(1)}(x-y)$ are the zeroth- and first-order terms in powers of $B$
in the translation invariant part of the propagator. [For the explicit forms of their Fourier transforms see 
Eqs.~(\ref{free-term}) and (\ref{linear-term}) above.] Of course, the representations in 
Eqs.~(\ref{free-propagator-magnetic-1}) and (\ref{free-propagator-magnetic-2}) are equivalent. One can 
check this explicitly, for example, by making use of the Landau gauge for the external field 
$A^{\rm ext}_{\nu}$.

Furthermore, Eq.~({\ref{self-energy}) implies that a similar expansion takes place also for the fermion self-energy
\begin{equation}
\Sigma(u,v)= \bar{\Sigma}^{(0)}(u-v)+i\Phi(u,v)\bar{\Sigma}^{(0)}(u-v)+ \bar\Sigma^{(1)}(u-v).
\end{equation}
The Fourier transforms of the self-energies $\bar{\Sigma}^{(0)}(x-y)$ and $\bar{\Sigma}^{(1)}(x-y)$ are given by
Eqs.~(\ref{self-energy-momentum-space-B0}) and (\ref{self-energy-momentum-space-B1}), respectively.

Omitting the noninteresting zeroth order in $B$ contribution in Eq.~(\ref{gxy-expansion}), we arrive at the following 
linear in $B$ contribution to the propagator: 
\begin{eqnarray}
G^{(1)}(x,x)&=&\bar{S}^{(1)}(x,x) 
+i\int  d^4 u d^4 v\left[\bar{S}^{(1)}(x-u)\bar{\Sigma}^{(0)}(u-v)\bar{S}^{(0)}(v-x)+\bar{S}^{(0)}(x-u)\bar{\Sigma}^{(0)}(u-v) 
\bar{S}^{(1)}(v-x)\right]\nonumber\\
&+&i\int  d^4 u d^4 v \left[\bar{S}^{(0)}(x-u)\bar\Sigma^{(1)}(u-v)\bar{S}^{(0)}(v-x) \right]\nonumber\\
&-&\int  d^4 u d^4 v \left[\Phi(x,u)+\Phi(u,v)+\Phi(v,x)\right]\bar{S}^{(0)}(x-u)\bar{\Sigma}^{(0)}(u-v)\bar{S}^{(0)}(v-x).
\label{Mat-propagator-linear-magnetic}
\end{eqnarray}
Noting that $\Phi(x,u)+\Phi(u,v)+\Phi(v,x) = -\frac{eB}{2}\left[(x_1-u_1)(v_2-x_2)-(v_1-x_1) (x_2-u_2)\right]$ 
is a translation invariant function, it is convenient to switch to the momentum space on the right-hand 
side of Eq.~(\ref{Mat-propagator-linear-magnetic}). The result reads 
\begin{eqnarray}
G^{(1)}(x,x)&=& \int \frac{d^4 p}{(2\pi)^4}\bar{S}^{(1)}(p)
+i\int \frac{d^4 p}{(2\pi)^4}\left[ 
\bar{S}^{(1)}(p)\bar{\Sigma}^{(0)}(p)\bar{S}^{(0)}(p) 
+\bar{S}^{(0)}(p)\bar{\Sigma}^{(0)}(p)\bar{S}^{(1)}(p) 
+\bar{S}^{(0)}(p)\bar\Sigma^{(1)}(p)\bar{S}^{(0)}(p) \right]\nonumber\\
&-&\frac{eB}{2}\int\frac{d^4 p}{(2\pi)^4}\left[\frac{\partial\bar{S}^{(0)}(p) }{\partial p_1}
\bar{\Sigma}^{(0)}(p)\frac{\partial\bar{S}^{(0)}(p) }{\partial p_2}
-\frac{\partial\bar{S}^{(0)}(p)}{\partial p_2}\bar{\Sigma}^{(0)}(p)\frac{\partial\bar{S}^{(0)}(p) }{\partial p_1} \right].
\label{linear-propagator}
\end{eqnarray}
By substituting this into the definition in Eq.~(\ref{j53-general}), we obtain the following expression 
for the axial current density:
\begin{eqnarray}
\langle j_{5}^{3}\rangle=\langle j_{5}^{3}\rangle_0+\langle j_{5}^{3}\rangle_{\alpha},
\label{axial-current-complete}
\end{eqnarray}
where
\begin{eqnarray}
\langle j_{5}^{3}\rangle_0= - \int \frac{d^4 p}{(2\pi)^4}\mbox{tr} \left[ \gamma^3\gamma^5 \bar{S}^{(1)}(p) \right]
\end{eqnarray}
is the contribution to the axial current in the free theory and
\begin{eqnarray}
\langle j_{5}^{3}\rangle_{\alpha} &=&
\frac{eB}{2} \int \frac{d^4 p}{(2\pi)^4}\mbox{tr} \Bigg[ \gamma^3\gamma^5 
\frac{\partial \bar{S}^{(0)}(p) }{\partial p_1}\bar{\Sigma}^{(0)}(p)  \frac{\partial \bar{S}^{(0)}(p) }{\partial p_2}
- \gamma^3\gamma^5  \frac{\partial \bar{S}^{(0)}(p) }{\partial p_2}\bar{\Sigma}^{(0)}(p)\frac{\partial \bar{S}^{(0)}
(p)}{\partial p_1} \Bigg]
\nonumber\\
&-&i\int \frac{d^4 p}{(2\pi)^4} \mbox{tr} \Bigg[ 
\gamma^3\gamma^5 \bar{S}^{(1)}(p)\bar{\Sigma}^{(0)}(p)\bar{S}^{(0)}(p) 
+\gamma^3\gamma^5 \bar{S}^{(0)}(p)\bar{\Sigma}^{(0)}(p) \bar{S}^{(1)}(p) 
+ \gamma^3\gamma^5 \bar{S}^{(0)}(p) \bar{\Sigma}^{(1)}(p)\bar{S}^{(0)}(p) \Bigg]
+\langle j_{5}^{3}\rangle_{\rm ct}\nonumber\\
\label{axial-current-initial}
\end{eqnarray}
defines the leading radiative corrections to the axial current. The counterterm contribution $\langle j_{5}^{3}\rangle_{\rm ct}$ in
Eq.~(\ref{axial-current-initial}) contains all the contributions with $\delta_2$ and $\delta_m$. 
Its explicit form will be given in Sec.~\ref{Sec:Counterterms} below.

It is instructive to start from investigating the structure of Eq.~(\ref{axial-current-complete}) in the 
free theory (i.e., to the zeroth order in $\alpha$). By making use of the explicit form of 
$\bar{S}^{(1)}(k) $ in Eq.~(\ref{linear-term}), we straightforwardly derive the following 
contribution to the axial current density:
\begin{equation}
\langle j_{5}^{3}\rangle_0
=-\frac{eB\,\mbox{sign}(\mu)}{4\pi^3}\int d^3\mathbf{k}\,\delta(\mu^2-\mathbf{k}^2-m^2)
=-\frac{eB\,\mbox{sign}(\mu)}{2\pi^2}\sqrt{\mu^2-m^2},\,
\label{axial-current-topological}
\end{equation}
which coincides, of course, with the very well-known topological contribution \cite{Zhitnitsky}. 
Note that in contrast to the approach using the expansion over the Landau levels, where the 
contribution to $\langle j_{5}^{3}\rangle_0$ comes only from the filled LLL states, the origin 
of the same topological contribution in the formalism of weak magnetic fields is quite different.
As Eq.~(\ref{axial-current-topological}) implies, it comes from the Fermi surface and, therefore,
provides a dual description of the topological contribution in this formalism. (Interestingly, 
the origin of the topological contribution in the weak field analysis above may have some similarities with 
the Wigner function formalism \cite{Gao:2012ix}.)

By substituting the propagators (\ref{free-term}) and (\ref{linear-term}) into Eq.~(\ref{axial-current-initial}), 
we find the following leading radiative corrections to the axial current:
\begin{eqnarray}
\langle j_{5}^{3}\rangle_{\alpha} &=& 32 \pi \alpha eB \int\frac{d^4 p\,d^4 k}{(2\pi)^8}  \frac{1}{(P-K)^2_{\Lambda}}
\Bigg[\frac{(k_0+\mu)  [(p_0+\mu)^2+p_\perp^2- p_3^2-m^2 ] -2(p_0+\mu)(p_1k_1+p_2k_2) }
{(P^2-m^2)^3 (K^2-m^2)}
\nonumber\\
&&-2\frac{(p_0+\mu)  (p_1k_1+p_2k_2+2k_3p_3+4m^2)-(k_0+\mu)[(p_0+\mu)^2+p_3^2+m^2]}
{(P^2-m^2)^3 (K^2-m^2)}
\nonumber\\
&&-\frac{(k_0+\mu) [(p_0+\mu)^2-p_\perp^2+ p_3^2+m^2] - 2 (p_0+\mu) p_3k_3}{(P^2-m^2)^2 
(K^2-m^2)^2}
\Bigg]+\langle j_{5}^{3}\rangle_{\rm ct}\nonumber\\
&=&32 \pi \alpha eB \int\frac{d^4 p\,d^4 k}{(2\pi)^8}  \frac{1}{(P-K)^2_{\Lambda}}
\Bigg[\frac{(k_0+\mu)  [3(p_0+\mu)^2+\mathbf{p}^2+m^2 ] -4(p_0+\mu)(\mathbf{p}\cdot\mathbf{k}
+2m^2) }{(P^2-m^2)^3 (K^2-m^2)}
\nonumber\\
&& - \frac{(k_0+\mu) [3(p_0+\mu)^2-\mathbf{p}^2+3m^2] - 2 (p_0+\mu) (\mathbf{p}\cdot \mathbf
{k})}
{3(P^2-m^2)^2 (K^2-m^2)^2}
\Bigg]+\langle j_{5}^{3}\rangle_{\rm ct}.
\label{j53-1}
\end{eqnarray}
Here we use the shorthand notation $K^2=[k_0+\mu+i\epsilon\, \mbox{sign}(k_0)]^2-\mathbf{k}^2$
and $P^2=[p_0+\mu+i\epsilon\, \mbox{sign}(p_0)]^2-\mathbf{p}^2$. As for the definition 
of $(P-K)^2_{\Lambda}$, it follows Eq.~(\ref{photon-propagator}). Furthermore, the following replacements 
have been made in the integrand: $p_\perp^2\to \frac{2}{3}\mathbf{p}^2$, $p_3^2\to \frac{1}{3}\mathbf{p}^2$, 
and $p_3k_3 \to \frac{1}{3}(\mathbf{p}\cdot \mathbf{k})$. These replacements are allowed by the 
rotational symmetry of the other parts of the integrand. 

\subsection{Integration by parts}
\label{Sec:Integration}

It is convenient to represent Eq.~(\ref{j53-1}) as follows:
\begin{eqnarray}
\langle j_{5}^{3}\rangle_{\alpha}
&=&32 \pi \alpha eB \int\frac{d^4 p d^4 k}{(2\pi)^8}  \frac{1}{(P-K)^2_{\Lambda}}
\Bigg[\frac{4(p_0+\mu)[(k_0+\mu) (p_0+\mu)  -\mathbf{p}\cdot\mathbf{k}-2m^2] }{(P^2-m^2)^3 (K^2-m^2)}
-\frac{(k_0+\mu)}{(P^2-m^2)^2 (K^2-m^2)}
\nonumber\\
&& - \frac{(k_0+\mu) [3(p_0+\mu)^2-\mathbf{p}^2+3m^2- 2 (\mathbf{p}\cdot \mathbf{k})]}
{3(P^2-m^2)^2 (K^2-m^2)^2}\Bigg]+\langle j_{5}^{3}\rangle_{\rm ct}.
\label{j53-2}
\end{eqnarray}
Since the denominators of the integrand in this expression contain the factors $(P^2-m^2)^n$ and $(K^2-m^2)^n$, 
with $n = 2, 3$, which vanish on the Fermi surface, the integrand in (\ref{j53-2}) is singular there. Therefore, 
one should carefully treat the singularities in the calculation of the axial current. For this, we find it very 
convenient to use the following identity valid for all integers $n\geq 1$:
\begin{eqnarray}
\frac{1}{ \left[ [k_0+\mu +i\epsilon\, \mbox{sign}(k_0)]^2-\mathbf{k}^2-m^2 \right]^n}&=&
 \frac{1}{\left[(k_0+\mu )^2-\mathbf{k}^2-m^2+i\epsilon \right]^n}\nonumber\\
&+&\frac{2\pi i(-1)^{n-1}}{(n-1)!}\theta(|\mu |-|k_0|) \theta(-k_0 \mu )\delta^{(n-1)}\left[(k_0+\mu )^2-\mathbf{k}^2-m^2\right],
\label{1overK^2-n}
\end{eqnarray}
which can be obtained from Eq.~(\ref{pole-identity}) in Appendix \ref{App:proper-time-rep}  
by differentiating it $n-1$ times with respect to $m^2$. Since the first term on the right-hand side has 
the pole prescription as in the theory without the filled fermion states, we call it the ``vacuum" part. 
The second term in this expression takes care of the filled fermion states, and we call it the ``matter"
part.

One can also obtain another useful relation by differentiating Eq.~(\ref{1overK^2-n}) 
with respect to energy $k_0$,
\begin{eqnarray}
&& \frac{\partial}{\partial k_0}\left(
\frac{1}{ \left[ [k_0+\mu+i\epsilon\, \mbox{sign}(k_0)]^2-m^2-\mathbf{k}^2 \right]^{n}} \right) 
= -\frac{2n(k_0+\mu)}{ \left[ [k_0+\mu+i\epsilon\, \mbox{sign}(k_0)]^2-m^2-\mathbf{k}^2 \right]^{n+1}} \nonumber \\
&&\hspace{2.5in}+ \frac{2\pi i (-1)^n \mbox{sign}(\mu)}{(n-1)!} \delta^{(n-1)} \left[(k_0+\mu)^2-\mathbf{k}^2-m^2\right]
\left[\delta(k_0) - \delta(k_0+\mu) \right],
\label{1overK^2-n-der}
\end{eqnarray}
where we made use of Eq.~(\ref{1overK^2-n}) the second time, albeit with $n\to n+1$, in order to render 
the result on the right-hand side in the form of the $(n+1)$th order pole with the conventional 
$i\epsilon$ prescription at nonzero $\mu$. In addition, we used the following easy to derive result:
\begin{equation}
\frac{\partial}{\partial k_0}\left[\theta(|\mu |-|k_0|) \theta(-k_0 \mu )\right] = \mbox{sign}(\mu)\left[\delta(k_0+\mu) - \delta(k_0) \right].
\end{equation}
We note that $\delta(k_0+\mu)$ in the last term on the right-hand side of Eq.~(\ref{1overK^2-n-der}) 
never contributes. Indeed, this $\delta$ function is nonvanishing only when $k_0+\mu=0$. It multiplies, however, 
another $\delta$ function, which is nonvanishing only when $(k_0+\mu)^2-\mathbf{k}^2-m^2=0$. Since the two 
conditions cannot be simultaneously satisfied, the corresponding contribution is trivial. After taking this into account, 
we finally obtain
\begin{eqnarray}
\frac{\partial}{\partial k_0}\left(
\frac{1}{ \left[ [k_0+\mu+i\epsilon\, \mbox{sign}(k_0)]^2-m^2-\mathbf{k}^2 \right]^{n}} 
\right)&=&-\frac{2n(k_0+\mu)}{ \left[ [k_0+\mu+i\epsilon\, \mbox{sign}(k_0)]^2-m^2-\mathbf{k}^2 \right]^{n+1}} \nonumber \\
&+& \frac{2\pi i (-1)^n \mbox{sign}(\mu)}{(n-1)!} \delta^{(n-1)} \left(\mu^2-\mathbf{k}^2-m^2\right) \delta(k_0).
\label{1overK^2-n-derivative}
\end{eqnarray}
Now, by making use of the above identities, we can proceed to the calculation of $\langle j_{5}^{3}\rangle_{\alpha}$ 
in Eq.~(\ref{j53-2}). We start by simplifying the corresponding expression using integrations by parts. 
Note that the $\Lambda$-regulated representation has nice convergence properties in the ultraviolet and, 
therefore, all integrations by parts in the analysis that follows will be perfectly justified.

The first term in $\langle j_{5}^{3}\rangle_{\alpha}$ in Eq.~(\ref{j53-2}) is proportional to $p_0+\mu$ 
and contains $(P^2-m^2)^3$ in the denominator. Therefore, we use identity (\ref{1overK^2-n-derivative}) 
with $n=2$ and $k\to p$, i.e.,
\begin{equation}
\frac{4(p_0+\mu)}{\left(P^2-m^2\right)^3}
=-\frac{\partial}{\partial p_0}\left(\frac{1}{(P^2-m^2)^2}\right)
+2i\pi \delta^{\prime}\left[\mu^2-m^2-\mathbf{p}^2\right]\delta(p_0).
\label{Mat-identity}
\end{equation}
Using it, we rewrite the first term in the integrand of Eq.~(\ref{j53-2}) as follows:
\begin{eqnarray}
\mbox{1st}&=& f_1 + 32 \alpha eB\int\frac{d^4 p d^4 k}{(2\pi)^8}  
\frac{ (k_0+\mu) (p_0+\mu)  -\mathbf{p}\cdot\mathbf{k}-2m^2 }{(P-K)^2_{\Lambda}(K^2-m^2)}
\frac{\partial }{\partial p_0}\left(\frac{-1}{(P^2-m^2)^2}\right) \nonumber\\
&=&f_1 + 32 \alpha eB\int\frac{d^4 p d^4 k}{(2\pi)^8}  \frac{1}{(P^2-m^2)^2}\frac{\partial }{\partial p_0}
\left(\frac{ (k_0+\mu) (p_0+\mu)  -\mathbf{p}\cdot\mathbf{k}-2m^2 }{(P-K)^2_{\Lambda}(K^2-m^2)}\right) 
\nonumber\\
&=&f_1 + 32 \alpha eB\int\frac{d^4 p d^4 k}{(2\pi)^8}  \frac{1}{(P^2-m^2)^2}
\left(\frac{ (k_0+\mu)}{(P-K)^2_{\Lambda}(K^2-m^2)}
+\frac{ (k_0+\mu)(p_0+\mu)  -\mathbf{p}\cdot\mathbf{k}-2m^2 }{(K^2-m^2)}\frac{\partial }{\partial p_0}\frac{ 1}{(P-K)^2_{\Lambda}}
\right),\nonumber\\
\label{Mat-first-term}
\end{eqnarray}
where the singular ``matter" term, containing the derivative of a $\delta$ function at the Fermi surface, 
was separated into a new function, 
\begin{equation}
f_1= 64 i \pi^2 \alpha eB  \int\frac{d^4 p d^4 k}{(2\pi)^8}  
\frac{ (k_0+\mu) (p_0+\mu) -\mathbf{p}\cdot\mathbf{k}-2m^2 }{(P-K)^2_{\Lambda}(K^2-m^2)}
\delta^{\prime}\left[\mu^2-m^2-\mathbf{p}^2\right]\delta(p_0).
\label{Mat-f-1}
\end{equation}
We note that the first term in the parentheses in Eq.~(\ref{Mat-first-term}) cancels with the second 
term in the integrand of Eq.~(\ref{j53-2}). Then using 
\begin{equation}
\frac{\partial}{\partial p_0}\frac{ 1}{(P-K)^2_{\Lambda}}=-\frac{\partial }{\partial k_0}\frac{ 1}{(P-K)^2_{\Lambda}}
\end{equation}
and integrating by parts, we find that the sum of the first and second terms in the integrand of Eq.~(\ref{j53-2}) is equal to
\begin{eqnarray}
(\mbox{1st}+\mbox{2nd})&=&f_1 + 32 \alpha eB \int\frac{d^4 p d^4 k}{(2\pi)^8}  \frac{ 1}{(P-K)^2_{\Lambda}(P^2-m^2)^2}
\left(
\frac{p_0+\mu}{(K^2-m^2)}
+\left[(k_0+\mu)(p_0+\mu)  -\mathbf{p}\cdot\mathbf{k}-2m^2\right]\frac{\partial }{\partial k_0}
\frac{ 1 }{(K^2-m^2)}
\right)
\nonumber\\
&=&f_1+f_2+32 \alpha eB \int\frac{d^4 p d^4 k}{(2\pi)^8}  \frac{ 1}{(P-K)^2_{\Lambda}}
\left( \frac{(p_0+\mu) }{(P^2-m^2)^2(K^2-m^2)}
-2(k_0+\mu)\frac{(k_0+\mu)(p_0+\mu)  -\mathbf{p}\cdot\mathbf{k}-2m^2}{(P^2-m^2)^2(K^2-m^2)^2} \right).\nonumber\\
\label{Mat-two-terms}
\end{eqnarray}
Note that here we used the identity
\begin{equation}
\frac{\partial}{\partial k_0}\left(\frac{1}{K^2-m^2}\right)
=\frac{-2(k_0+\mu)}{\left(K^2-m^2\right)^2}-2i\pi \delta \left(\mu^2-m^2-\mathbf{k}^2\right)\delta(k_0),
\label{Mat-derivative}
\end{equation}
which follows from Eq.~(\ref{1overK^2-n-derivative}) with $n=1$, and introduced another 
function, which contains the leftover contribution with the $\delta$ function, 
\begin{equation}
f_2=-64 i\pi^2 \alpha eB \int\frac{d^4 p d^4 k}{(2\pi)^8} \frac{(k_0+\mu)(p_0+\mu)  
-\mathbf{p}\cdot\mathbf{k}-2m^2}{(P-K)^2_{\Lambda}(P^2-m^2)^2}
\delta \left(\mu^2-m^2-\mathbf{k}^2\right)\delta(k_0).
\label{Mat-f-2}
\end{equation}
It is convenient to make the change of variables $p \to k$ and $k \to p$ in the first term in Eq.~(\ref{Mat-two-terms}). 
Then, the two terms in the integrand can be combined, resulting in
\begin{equation}
(\mbox{1st}+\mbox{2nd}) = f_1 +f_2+32 \alpha eB\int\frac{d^4 p d^4 k}{(2\pi)^8}  \frac{(k_0+\mu)
\left[-(p_0+\mu)^2  -\mathbf{p}^2+2\mathbf{p}\cdot\mathbf{k}+3m^2\right]}{(P-K)^2_{\Lambda}(P^2-m^2)^2(K^2-m^2)^2}.
\label{two-terms-1}
\end{equation}
Finally, by combining the result in Eq.~(\ref{two-terms-1}) with the last term in the integrand of Eq.~(\ref{j53-2}), we obtain
\begin{equation}
\langle j_{5}^{3}\rangle_{\alpha} =  f_1 +f_2 -\frac{64}{3} \pi \alpha eB \int\frac{d^4 p d^4 k}{(2\pi)^8}
\frac{(k_0+\mu) }{(P-K)_{\Lambda}^2}
\frac{3(P^2-m^2)+ 4\mathbf{p}\cdot(\mathbf{p}-\mathbf{k})}{(P^2-m^2)^2(K^2-m^2)^2}+\langle j_{5}^{3}\rangle_{\rm ct}.
\end{equation}
Using the identity in Eq.~(\ref{Mat-derivative}) once again, we rewrite the last expression as follows:
\begin{equation}
\langle j_{5}^{3}\rangle_{\alpha} =  f_1 +f_2+f_3  +\langle j_{5}^{3}\rangle_{\rm ct}
+\frac{64}{3} \pi \alpha eB \int\frac{d^4 p d^4 k}{(2\pi)^8} 
\frac{(k_0-p_0)}{(P-K)_{\Lambda}^4}\left(\frac{3}{(P^2-m^2)(K^2-m^2)}
+\frac{4\mathbf{p}\cdot(\mathbf{p}-\mathbf{k})}{(P^2-m^2)^2(K^2-m^2)}  
\right),
\label{j53-3}
\end{equation}
where
\begin{equation}
f_3 = \frac{ 64i \pi^2 \alpha eB }{3}\int\frac{d^4 p d^4 k}{(2\pi)^8}  
\frac{3(P^2-m^2)+ 4\mathbf{p}\cdot(\mathbf{p}-\mathbf{k})}{(P-K)_{\Lambda}^2(P^2-m^2)^2}
\delta \left(\mu^2-m^2-\mathbf{k}^2\right)\delta(k_0) .
\label{Mat-f-3}
\end{equation}
Since the first term of the integrand in Eq.~(\ref{j53-3}) is odd under the exchange 
$p \leftrightarrow k$, its contribution vanishes, and we obtain
\begin{equation}
\langle j_{5}^{3}\rangle_{\alpha} = f_1 +f_2+f_3+\langle j_{5}^{3}\rangle_{\rm ct}
+\frac{64}{3} \pi \alpha eB \int\frac{d^4 p d^4 k}{(2\pi)^8} \frac{(k_0-p_0)}{(P-K)_{\Lambda}^4} 
\frac{4\mathbf{p}\cdot(\mathbf{p}-\mathbf{k})}{(P^2-m^2)^2(K^2-m^2)}.
\end{equation}
Finally, by making use of the identity 
\begin{equation}
\frac{\mathbf{p}\cdot(\mathbf{p}-\mathbf{k})}{(P-K)_{\Lambda}^4} 
=\frac{1}{2}\mathbf{p}\cdot\bm{\nabla}_\mathbf{k}\frac{-1}{(P-K)_{\Lambda}^2}
\end{equation}
and integrating by parts, we derive
\begin{eqnarray}
\langle j_{5}^{3}\rangle_{\alpha}
&=& f_1 +f_2+f_3+\langle j_{5}^{3}\rangle_{\rm ct}+\frac{64}{3} \pi \alpha eB \int\frac{d^4 p d^4 k}{(2\pi)^8} 
\frac{2(k_0-p_0)}{(P^2-m^2)^2(K^2-m^2)}  
\mathbf{p}\cdot\bm{\nabla}_\mathbf{k}\frac{-1}{(P-K)_{\Lambda}^2}
\nonumber\\
&=& f_1 +f_2+f_3+\langle j_{5}^{3}\rangle_{\rm ct}+\frac{64}{3} \pi \alpha eB \int\frac{d^4 p d^4 k}{(2\pi)^8} 
\frac{2(k_0-p_0)}{(P-K)_{\Lambda}^2(P^2-m^2)^2}  
\mathbf{p}\cdot\bm{\nabla}_\mathbf{k}\frac{1}{(K^2-m^2)}
\nonumber\\
&=& f_1 +f_2+f_3 +\langle j_{5}^{3}\rangle_{\rm ct}+\frac{64}{3} \pi \alpha eB \int\frac{d^4 p d^4 k}{(2\pi)^8} 
\frac{4(k_0-p_0)\mathbf{p}\cdot\mathbf{k}}{(P-K)_{\Lambda}^2(P^2-m^2)^2(K^2-m^2)^2}  = f_1 +f_2+f_3 +\langle j_{5}^{3}\rangle_{\rm ct},
\label{chiral-current-new} 
\end{eqnarray}
where the last integral term in the last line of Eq.~(\ref{chiral-current-new}) vanishes 
because it is odd under the exchange $p \leftrightarrow k$. Collecting together all contributions, 
i.e., $f_1$ in Eq.~(\ref{Mat-f-1}), $f_2$ in Eq.~(\ref{Mat-f-2}) and $f_3$ in Eq.~(\ref{Mat-f-3}),
we have the following leading radiative corrections to the axial current:
\begin{eqnarray}
\langle j_{5}^{3}\rangle_{\alpha}&=&64i\pi^2\alpha eB \int\frac{d^4 p d^4 k}{(2\pi)^8}\Bigg[ 
\frac{ (k_0+\mu) (p_0+\mu) -\mathbf{p}\cdot\mathbf{k}-2m^2 }{(P-K)^2_{\Lambda}(K^2-m^2)}
\delta^{\prime}\left[\mu^2-m^2-\mathbf{p}^2\right]\delta(p_0) \nonumber\\
&+& \frac{3(p_0+\mu)^2-3(k_0+\mu)(p_0+\mu)+\mathbf{p}^2-\mathbf{p}\cdot\mathbf{k}+3m^2}{3(P-K)^2_{\Lambda}(P^2-m^2)^2}
\delta \left(\mu^2-m^2-\mathbf{k}^2\right)\delta(k_0)\Bigg]+\langle j_{5}^{3}\rangle_{\rm ct},
\label{j53-by-parts}
\end{eqnarray}
where the first term in the integrand comes from $f_1$, while the second term comes from 
the sum $f_2+f_3$. The result in Eq.~(\ref{j53-by-parts}) is quite remarkable for several reasons. 
From a technical viewpoint, it reveals that the integration by parts allowed us to reduce the original 
two-loop expression in Eq.~(\ref{j53-2}) down to a much simpler one-loop form. Indeed, after the 
integration over one of the momenta in Eq.~(\ref{j53-by-parts}) is performed using the $\delta$ functions 
in the integrand, the expression will have an explicit one-loop form. Such a simplification will turn out to 
be extremely valuable, allowing us to obtain an analytic result for the leading radiative corrections to 
the axial current.

In addition, the result in Eq.~(\ref{j53-by-parts}) reveals important physics details
about the origin of the radiative corrections to the axial current. It shows that all nonzero corrections
come from the regions of the phase space, where either $p$ or $k$ momentum is restricted 
to the Fermi surface. This resembles the origin of the topological contribution in 
Eq.~(\ref{axial-current-topological}). In both cases, the presence of the singular ``matter" 
terms in identities like (\ref{Mat-identity}) and (\ref{Mat-derivative}) was crucial for 
obtaining a nonzero result. Moreover, by tracing back the derivation of the result in 
Eq.~(\ref{j53-by-parts}), we see that all nonsingular terms are gone after the 
integration by parts. This makes us conclude that the nonzero radiative corrections 
to the axial current are intimately connected with the precise form of the singularities in the fermion 
propagator at the Fermi surface, that separates the filled fermion states with energies 
less than $\mu$ and empty states with larger energies.

\subsection{Counterterm contribution}
\label{Sec:Counterterms}

The calculation of the axial current in Eq.~(\ref{j53-by-parts}) is still technically quite involved. 
However, it is relatively straightforward to show [see also the derivation of 
Eq.~(\ref{f_1_UV-divergency}) in Appendix~\ref{App:calculation-j35}] that 
the right-hand side in (\ref{j53-by-parts}) without the counterterm has a logarithmically 
divergent contribution when $\Lambda\to \infty$, i.e., 
\begin{equation}
\frac{\alpha eB(2\mu^2+m^2)}
{4 \pi^3 \sqrt{\mu^2-m^2}}\ln\frac{\Lambda}{m}.
\label{j53-loop-UV}
\end{equation}
To cancel this divergence, we should add the contribution due the
counterterms in Lagrangian (\ref{Lagrangian}). The Fourier transform
of the translational invariant part of the counterterm contribution to the self-energy reads
\begin{equation}
\bar{\Sigma}^{(0)}_{\rm ct}(p)= \delta_2[(p_0+\mu)\gamma^0-\mathbf{p}\cdot\bm{\gamma}] -  \delta_m,
\end{equation}
where $\delta_2$ was defined in Eq.~(\ref{wave-function-renormalization-constant}), while 
$\delta_m = Z_2 m_0-m \simeq m\delta_2-\delta m$ and $\delta m$ was defined in 
Eq.~(\ref{mass-renormalization-constant}).

We find the following leading order contributions to the axial current density due to counterterms:
\begin{eqnarray}
\langle j_{5}^{3}\rangle_{\rm ct} &=&
-\delta_2\langle j_{5}^{3}\rangle_0
-4ieB\int\frac{d^4 p}{(2\pi)^4}\frac{\delta_2(p_0+\mu)}{(P^2-m^2)^2}
-8ieB\int\frac{d^4 p}{(2\pi)^4}\frac{(p_0+\mu) \left[ \delta_2 ( (p_0+\mu)^2-\mathbf{p}^2+m^2 ) -2 m\delta_m \right]}{(P^2-m^2)^3}
\nonumber\\
&=&
-8ieB\int\frac{d^4 p}{(2\pi)^4}\frac{(p_0+\mu) \left[ \delta_2 ( P^2-m^2) +2 m(m\delta_2-\delta_m) \right]}{(P^2-m^2)^3}
\nonumber\\
&=&
-8ieB\delta_2 \int\frac{d^4 p}{(2\pi)^4}\frac{p_0+\mu}{(P^2-m^2)^2}
-8 i m \left(m\delta_2-\delta_m\right)eB\frac{\partial }{\partial (m^2)}\int\frac{d^4 p}{(2\pi)^4}\frac{p_0+\mu}{(P^2-m^2)^2}
\nonumber\\
&=&-\frac{eB}{\pi^2}\sqrt{\mu^2-m^2}\delta_2+\frac{eB m \left(m\delta_2-\delta_m\right) }{2\pi^2\sqrt{\mu^2-m^2}}.
\label{j35-ct}
\end{eqnarray}
Here we used the same result of integration as in the topological term, see Eq.~(\ref{axial-current-topological}).

By making use of the explicit form of the counterterms (\ref{wave-function-renormalization-constant}) and 
(\ref{mass-renormalization-constant}), we obtain
\begin{equation}
\langle j_{5}^{3}\rangle_{\rm ct} =
-\frac{\alpha eB}{2\pi^3}\sqrt{\mu^2-m^2}\left(\frac{1}{2}\ln\frac{\Lambda^2}{m^2}+\ln\frac{m_\gamma^2}{m^2}+\frac{9}{4}\right)  
-\frac{3\alpha eB m^2}{4\pi^3\sqrt{\mu^2-m^2}}\left(\frac{1}{2}\ln\frac{\Lambda^2}{m^2} +\frac{1}{4}\right).
\label{counterterm-contribution-general}
\end{equation} 
For $m\ll|\mu|$, it reduces to 
\begin{equation}
\langle j_{5}^{3}\rangle_{\rm ct} \simeq
-\frac{\alpha eB \mu}{2\pi^3} \left(\frac{1}{2}\ln\frac{\Lambda^2}{m^2}+\ln\frac{m_\gamma^2}{m^2}+\frac{9}{4}\right)  
-\frac{\alpha eB m^2}{2\pi^3\mu}\left(\frac{1}{2}\ln\frac{\Lambda^2}{m_\gamma^2} -\frac{3}{4}\right).
\label{counterterm-contribution-general-small-m}
\end{equation}

\subsection{The final result}

The complete expression for the leading radiative corrections to the axial current is given by 
Eq.~(\ref{chiral-current-new}). It consists of the counterterm contribution, calculated in the previous 
subsection, and the additional matter contribution $f_1 +f_2+f_3$. The latter is calculated 
in Appendix~\ref{App:calculation-j35}. For $m \ll |\mu|$, it reads
\begin{equation}
f_1+f_2+f_3 =  \frac{\alpha eB\mu}{2\pi^3} \left(\ln\frac{\Lambda}{2\mu}+\frac{11}{12}\right)
+\frac{\alpha eBm^2}{2 \pi^3 \mu} \left(\ln\frac{\Lambda}{2^{3/2}\mu}+\frac{1}{6}\right).
\end{equation}
Note that this expression has the right ultraviolet logarithmic divergencies (when $\Lambda\to\infty$)
that will cancel exactly with those in the counterterm (\ref{counterterm-contribution-general-small-m}). 
Combining the two results, we finally obtain the following leading 
radiative corrections to the axial current in the case $m \ll |\mu|$:
\begin{equation}
\langle j_{5}^{3}\rangle_{\alpha}=-\frac{\alpha eB \mu}{2\pi^3}\left(\ln\frac{2\mu}{m}+\ln\frac{m_\gamma^2}{m^2}+\frac{4}{3}\right)  
-\frac{\alpha eB m^2}{2\pi^3\mu} \left(\ln\frac{2^{3/2}\mu}{m_\gamma} -\frac{11}{12}\right).
\label{j53-small-m-final}
\end{equation}
As expected, this result is independent of the ultraviolet regulator $\Lambda$. It does contain,
however, the dependence on the fictitious photon mass $m_\gamma$. This is the only infrared 
regulator left in our result. Its origin can be easily traced back to the infrared singularity
of the wave function renormalization $Z_2$ in the Feynman gauge used. As we discuss 
in the next section, this singularity is typical for a class of QED observables, obtained by 
perturbative methods. As we will explain below, in the complete physical expression for 
the axial current, obtained by going beyond the simplest double expansion in the coupling 
constant and magnetic field, the regulator $m_\gamma^2$ will likely be replaced by a physical
scale, e.g., such as $|eB|$ or $\alpha\mu^2$.

\section{Discussions and Conclusions}
\label{Sec:Conclusion}

Our study of the chiral separation effect in dense QED in the limit of a weak magnetic field suggests 
a conceptually new way to interpret and calculate the axial current density even in noninteracting 
theory. In contrast to the original formulation, which suggests that the topological contribution 
comes exclusively from the LLL filled states \cite{Zhitnitsky}, we show that the origin of the same 
contribution in the formalism of weak magnetic fields (\ref{axial-current-topological}) is quite 
different: it comes from the whole Fermi surface. Such a dual description of the topological 
contribution is of interest on its own. It is sensible to suggest that the underlying 
origin for such a dual description must be connected with the topological nature of the effect. 
It remains to be sorted out how this happens in detail.

Our result for the axial current density obtained perturbatively in the coupling constant and in 
linear order in the external magnetic field shows that the chiral separation effect in QED has 
nonvanishing radiative corrections. To leading order, these corrections are shown to be directly 
connected with the Fermi surface singularities in the fermion propagator at nonzero density. 
This interpretation is strongly supported by another observation: had we ignored the 
corresponding singular terms in the fermion propagator, the calculation of the two-loop 
radiative corrections would give a vanishing result.

The final result for the leading radiative corrections to the axial current density is presented 
in Eq.~(\ref{j53-small-m-final}). This is obtained by a direct calculation of all relevant
contributions to linear order in $\alpha$ and to linear order in the external magnetic field 
(strictly speaking, linear in $eB$ because the field always couples with the charge). 
The result in Eq.~(\ref{j53-small-m-final}) is presented in terms of renormalized (physical) 
parameters. As expected, it is independent of the ultraviolet regulator $\Lambda$, used 
at intermediate stages of calculations. This is a nontrivial statement since the original 
two-loop expression for the leading radiative corrections contains ultraviolet divergencies. 
In fact, the divergencies are unavoidable because the corresponding diagrams contain the 
insertions of the one-loop self-energy and vertex diagrams, which are known to have 
logarithmic divergencies. However, at the end of the day, all such divergencies 
are canceled exactly with the contributions due to the counterterms.

Our analysis shows that the matter contribution, $f_1+f_2+f_3$, to the axial current density 
(calculated in the Feynman gauge) has no additional singularities. While functions $f_1$ 
and $f_2+f_3$ separately do have additional infrared singularities, the physically relevant 
result for the sum $f_1+f_2+f_3$ is finite, see Appendix~\ref{App:calculation-j35} for details. 
As we see from Eq.~(\ref{j53-small-m-final}), however, the final result depends on the photon 
mass $m_\gamma$, which was introduced as the conventional infrared regulator. This 
feature deserves some additional discussion. 

It is straightforward to trace the origin of the $m_\gamma$ dependence in Eq.~(\ref{j53-small-m-final}) 
to the calculation of the well-known result for the wave function renormalization constant $\delta_2$, 
presented in Eq.~(\ref{wave-function-renormalization-constant}). In fact, this infrared problem 
is common for dynamics in external fields in QED (for a thorough discussion, see Sec.~14 in \cite{Weinberg}). 
The most famous example is provided by the calculation of the Lamb shift, when an electron is in a Coulomb
field. The point is that even for a light nucleus with $Z\alpha \ll 1$, one cannot consider the Coulomb field as a
weak perturbation in deep infrared. The reason is that this field essentially changes the dispersion relation for the electron 
at low energy and momenta. As a result, its four-momenta are not on the electron mass shell, where the infrared divergence
is generated in the renormalization constant $Z_2$. Because of that, this infrared divergence is fictitious. The
correct approach is to consider the Coulomb interaction perturbatively only at high energies, while to treat it
nonperturbatively at low energies. The crucial point is matching those two regions that leads to replacing the
fictitious parameter $m_{\gamma}$ by a physical infrared scale. This is the main subtlety that makes the 
calculation of the Lamb shift quite involved \cite{Weinberg}.

In the case of the Lamb shift, the infrared scale is related to the atomic binding energy, or equivalently 
the inverse Bohr radius. For smaller energies and momenta, the electron wave functions cannot possibly 
be approximated with plane waves, which is the tacit assumption of the weak field approximation. 
Almost exactly the same line of arguments applies in the present problem of  QED in an external 
magnetic field. In particular, the fermion momenta perpendicular to the magnetic field cannot be 
defined with a precision better than $\sqrt{|eB|}$, or equivalently the inverse magnetic length. This
implies that the contribution to the axial current, which comes from the low-energy photon exchange 
between the fermion states near the Fermi surface, should be treated nonperturbatively. Just like in the 
Lamb shift problem \cite{Weinberg}, we can anticipate that a proper nonperturbative treatment will 
result in a term proportional to $\ln(|eB|/m_\gamma^2)$, with a coefficient such as to cancel the 
$m_\gamma$ dependence in Eq.~(\ref{j53-small-m-final}). 

The additional complication in the problem at hand, which is absent in the study of the Lamb shift, 
is a nonzero density of matter. While doing the expansion in $\alpha$ and keeping only the leading 
order corrections, we ignored all screening effects, which formally appear to be of higher order. 
It is understood, however, that such effects can be very important at nonzero density. In particular, 
they could replace the unphysical infrared regulator $m_\gamma^2$ with a physical screening 
mass, i.e., the Debye mass $\sqrt{\alpha}\mu$.

In contrast to the physics underlying the Lamb shift, where the nonperturbative result 
can be obtained with the logarithmic accuracy by simply replacing $m_\gamma$ with the only 
physically relevant infrared scale in the problem, the same is not possible in the problem of the 
axial current at hand. The major complication here comes from the existence of two different 
physical regulators that can replace the unphysical infrared scale $m_\gamma$. One of 
them is $\sqrt{|eB|}$ and the other is $\sqrt{\alpha}\mu$. Because of the use of a double 
perturbative expansion in the analysis controlled by the small parameters $|eB|/\mu^2$ 
and $\alpha$, it is not possible to unambiguously resolve (without performing a direct 
nonperturtative calculation) which one of the two scales (or their combination) will 
cure the singularity in Eq.~(\ref{j53-small-m-final}).

Another natural question to address is the chiral limit, $m \to 0$. As one can see from
Eq. (\ref{j53-small-m-final}),  the current $\langle j_{5}^{3}\rangle_{\alpha}$ is singular in this limit.
This point reflects the well-known fact that massless QED possesses new types of infrared singularities: 
beside the well-known divergences connected with soft photons, there are also divergences connected 
with the emission and absorption of collinear fermion-antifermion pairs \cite{Kinoshita:1962ur,Lee:1964is}. 
In addition, because of a Gaussian infrared fixed point in massless QED, the renormalized electric charge 
of massless fermions is completely shielded. One can show that this property is also intimately related to 
the collinear infrared divergences \cite{Fomin:1976am}. The complete screening of the renormalized 
electric charge makes this theory very different from massive QED. It remains to be examined whether 
there is a sensible way to describe the interactions with external electromagnetic fields in massless 
QED \cite{footnote}.

In addition to the quantitative study of the nonperturbative low-energy contributions and the effect of screening,
there remain several other interesting problems to investigate in the future. Here we will mention only the following 
three. (i) It is of special interest to clarify the connection of the nontrivial radiative corrections to the axial 
current density calculated in this paper with the generation of the chiral shift parameter in dense 
QED. The analysis in the recent Ref. \cite{Gorbar:2013uga} shows that there is indeed such a 
connection but it is more complicated than that in the NJL model \cite{chiral-shift-3,chiral-shift-1,chiral-shift-2}. 
(ii) In order to make a contact with the physics of heavy-ion collisions, it would be interesting
to generalize our study to the case of a nonzero temperature. The corresponding study in the NJL model 
\cite{chiral-shift-3} suggests that the temperature dependence of the axial current density should be weak.  
(iii) The analysis made in the NJL model shows a lot of similarities between the structure of the 
axial current in the CSE effect \cite{chiral-shift-3,chiral-shift-1,chiral-shift-2} with that of the 
electromagnetic current in the CME one \cite{Fukushima:2010zza}. On the other hand, 
the arguments of Ref.~\cite{Rubakov:2010qi} may suggest that the dynamical 
part of the result for the electromagnetic current should vanish, while the topological contribution 
(which needs to be added as part of the modified conserved axial current) will have no radiative 
corrections. It remains to be seen if these expectations will be supported by direct calculations of 
the induced electromagnetic current in the CME effect in QED with a chiral chemical 
potential $\mu_5$.

\acknowledgments
We thank Gerald Dunne and Kenji Fukushima for useful remarks. I.A.S. would like to thank 
Andreas Schmitt and De-fu Hou for interesting discussions. The work of E.V.G. was supported 
partially by the European FP7 program, Grant No. SIMTECH 246937, and SFFR of Ukraine, 
Grant No.~F53.2/028. The work of V.A.M. was supported by the Natural Sciences and Engineering 
Research Council of Canada. He also acknowledges partial support by the National Science 
Foundation under Grant No.~PHYS-1066293 and the hospitality of the Aspen Center for Physics.
The work of I.A.S. and X.W. was supported in part by the U.S. National Science Foundation 
under Grant No.~PHY-0969844.

\appendix

\section{Schwinger parametrization for the fermion propagator at $B\neq 0$ and $\mu\neq 0$}
\label{App:proper-time-rep}

The proper-time representation for the fermion propagator in a constant external magnetic 
field was obtained long time ago by Schwinger \cite{Schwinger:1951nm}. A naive generalization 
of the corresponding representation to the case of a nonzero chemical potential (or density)
does not work however. This is due to the complications in the definition of the causal Feynman 
propagator in the complex energy plane when $\mu\neq 0$. The correct analytical 
properties of such a propagator describing particles above Fermi surface propagating forward 
in time and holes below Fermi surface propagating backward in time are implemented by 
introducing an appropriate $i\epsilon$ prescription. In particular, one replaces $k_0+\mu$ 
with $k_0+\mu+i\epsilon\, \mbox{sign}(k_0)$, where $\epsilon$ is a vanishingly small 
positive parameter.
For example, in the Landau level representation, the Fourier transform of the translation 
invariant part of the fermion propagator is defined as follows:
\begin{equation}
\bar{S}(k) = ie^{-k_{\perp}^2 \ell^{2}}\sum_{n=0}^{\infty}
\frac{(-1)^n D_{n}(k)}{ [k_0+\mu+i\epsilon\, \mbox{sign}(k_0)]^2 -m^2-k_{3}^2-2n|e B |} ,
\label{Fourier-tranlation-inv-S}
\end{equation}
where the residue at each individual Landau level is determined by
\begin{equation}
D_{n}(k) = 2\left[(k_0+\mu)\gamma^{0}+m-k^{3}\gamma^3\right]
\left[P_{-}L_n\left(2 k_{\perp}^2 \ell^{2}\right)
-P_{+}L_{n-1}\left(2 k_{\perp}^2 \ell^{2}\right)\right]
 + 4(\bm{k}_{\perp}\cdot\bm{\gamma}_{\perp}) L_{n-1}^1\left(2 k_{\perp}^2 \ell^{2}\right),
\end{equation}
where $L^{\alpha}_n(x)$ are associated Laguerre polynominals. 

Let us start by reminding the usual Schwinger's proper-time representation at zero fermion density, i.e., 
\begin{equation}
\frac{1}{[k_0+i\epsilon\, \mbox{sign}(k_0)]^2-{\cal M}_n^2} 
\equiv \frac{1}{k_0^2-{\cal M}_n^2+i\epsilon} 
= -i\int_{0}^{\infty} d s e^{is(k_0^2-{\cal M}_n^2+i\epsilon)} ,
\label{proper-time-concept}
\end{equation}
where ${\cal M}_n^2 = m^2+k_{3}^2+2n|e B |$. It is important to emphasize that the convergence 
of the integral and, thus, the validity of the representation 
are ensured by having the positive parameter $\epsilon$ in the exponent. Unfortunately, 
such a representation fails at finite fermion density. Indeed, by taking into account that 
\begin{equation}
\frac{1}{ [k_0+\mu+i\epsilon\, \mbox{sign}(k_0)]^2-{\cal M}_n^2} \equiv
\frac{1}{(k_0+\mu)^2-{\cal M}_n^2+i\epsilon\, \mbox{sign}(k_0)\mbox{sign}(k_0+\mu)},
\end{equation}
we see that the sign of the $i\epsilon$ term in the denominator is not fixed any more. The corresponding 
sign is determined by the product of $\mbox{sign}(k_0)$ and $\mbox{sign}(k_0+\mu)$ and 
can change, depending on the values of $k_0$ and $\mu$. For example, while it is positive
for $|k_0|>|\mu|$, it turns negative when $|k_0|<|\mu|$ and $k_0 \mu<0$. This seemingly 
innocuous property causes a serious problem for the integral representation utilized in 
Eq.~(\ref{proper-time-concept}). The sign changing $i\epsilon$ term in the exponent 
invalidates the representation at least for a range of quasiparticle energies.

In order to derive a modified proper-time representation for the fermion propagator, 
we will make use of the following identity:
\begin{eqnarray}
&&\frac{1}{ [k_0+\mu+i\epsilon\, \mbox{sign}(k_0)]^2-{\cal M}_n^2 }
=\frac{\theta(|k_0|-|\mu|)}{(k_0+\mu)^2-{\cal M}_n^2+i\epsilon}
+\theta(|\mu|-|k_0|) 
\left(\frac{\theta(k_0 \mu)}{(k_0+\mu)^2-{\cal M}_n^2+i\epsilon}
+\frac{\theta(-k_0 \mu)}{(k_0+\mu)^2-{\cal M}_n^2-i\epsilon} \right) \nonumber \\
&&\hspace{1in}= \frac{1}{(k_0+\mu)^2-{\cal M}_n^2+i\epsilon}
-\theta(|\mu|-|k_0|) \theta(-k_0 \mu)\left(
\frac{ 1 }{(k_0+\mu)^2-{\cal M}_n^2+i\epsilon}
-\frac{1}{(k_0+\mu)^2-{\cal M}_n^2-i\epsilon}
\right)
\nonumber \\
&&\hspace{1in}=  \frac{1}{(k_0+\mu)^2-{\cal M}_n^2+i\epsilon}
+2 i \pi \, \theta(|\mu|-|k_0|) \theta(-k_0 \mu)\delta \left[(k_0+\mu)^2-{\cal M}_n^2\right].
\label{pole-identity}
\end{eqnarray}
The first term on the right-hand side of 
Eq.~(\ref{pole-identity}) has a vacuumlike $i\epsilon$ prescription and, thus, allows a 
usual proper-time representation. The second term is singular and represents the 
additional ``matter" piece, which would be lost in the naive proper-time representation.
After making use of this identity, we derive the following modified proper-time representation 
for the propagator:
\begin{eqnarray}
\bar{S}(k) &=& e^{-k_{\perp}^2 \ell^{2}}
\sum_{n=0}^{\infty} (-1)^n D_{n}(k)
\int_{0}^{\infty} ds\, e^{i s [(k_0+\mu)^2-m^2-k_{3}^2-2n|e B|+i\epsilon]}\nonumber \\
&&-\theta(|\mu|-|k_0|)\theta(-k_0\mu) e^{-k_{\perp}^2 \ell^{2}}
\sum_{n=0}^{\infty} (-1)^n D_{n}(k)
\Bigg[
\int_{0}^{\infty} ds\, e^{i s [(k_0+\mu)^2-m^2-k_{3}^2-2n|e B|+i\epsilon]}\nonumber \\
&&+\int_{0}^{\infty} ds\, e^{-i s [(k_0+\mu)^2-m^2-k_{3}^2-2n|e B|-i\epsilon]}
\Bigg].
\end{eqnarray}
In order to perform the sum over the Landau levels, we use the following result for the 
infinite sum of the Laguerre polynominals:
\begin{equation}
\sum\limits_{n=0}^{\infty} z^n L_n^\alpha(x) = \frac{1}{(1-z)^{1+\alpha}}\exp\left(\frac{x z}{z-1}\right).
\label{table-sum}
\end{equation}
Then we obtain
\begin{eqnarray}
\bar{S}(k)&=& 
\int_{0}^{\infty} ds\, e^{i s [(k_0+\mu)^2-m^2-k_{3}^2+i\epsilon]-i k_{\perp}^2 \ell^{2} \tan(s |e B|)}
\left[(k_0+\mu)\gamma^{0}+m-\mathbf{k}\cdot\bm{\gamma}-(k^1\gamma^2-k^2\gamma^1)\tan(s e B)\right]
\nonumber \\
&&\times
\left[1+ \gamma^1\gamma^2 \tan(s e B) \right] 
-\theta(|\mu|-|k_0|) \theta(-k_0\mu) \nonumber \\
&\times &
\Bigg\{
\int_{0}^{\infty} ds 
    e^{i s [(k_0+\mu)^2-m^2-k_{3}^2+i\epsilon]-i k_{\perp}^2 \ell^{2} \tan(s |e B|)} 
\left[(k_0+\mu)\gamma^{0}+m-\mathbf{k}\cdot\bm{\gamma}-(k_1\gamma^2-k_2\gamma^1)\tan(s e B)\right]\nonumber \\
&&\times
\left[1+ \gamma^1\gamma^2 \tan(s e B) \right]
\nonumber \\
&&+\int_{0}^{\infty} ds 
    e^{- i s [(k_0+\mu)^2-m^2-k_{3}^2-i\epsilon]+i k_{\perp}^2 \ell^{2} \tan(s |e B|)}
    \left[(k_0+\mu)\gamma^{0}+m-\mathbf{k}\cdot\bm{\gamma}+(k_1\gamma^2-k_2\gamma^1)\tan(s e B)\right]\nonumber \\
&&\times
\left[1 - \gamma^1\gamma^2 \tan(s e B) \right]
\Bigg\}.
\label{S-prop-time-mu1}
\end{eqnarray}
This is a very convenient alternative representation for the fermion propagator in a constant 
external magnetic when $\mu\neq 0$. It allows, in particular, a straightforward derivation of the 
expansion in powers of the magnetic field. To zeroth order in magnetic field, we obtain
\begin{eqnarray}
\bar{S}^{(0)}(k) = \bar{S}^{(0)}_{\rm vac}(k)+\bar{S}^{(0)}_{\rm mat}(k),
\label{S0-vac-plus-matter}
\end{eqnarray}
where
\begin{eqnarray}
\bar{S}^{(0)}_{\rm vac}(k)= 
\int_{0}^{\infty} ds\, e^{i s [(k_0+\mu)^2-m^2-\mathbf{k}^2+i\epsilon]}\,\left[(k_0+\mu)\gamma^{0}+m-\mathbf{k}\cdot\bm{\gamma}\right]
\end{eqnarray}
and
\begin{eqnarray}
\bar{S}^{0)}_{\rm mat}(k)=-2\pi \, \theta(|\mu|-|k_0|) \theta(-k_0\mu)\,\left[(k_0+\mu)\gamma^{0}+m-\mathbf{k}\cdot\bm{\gamma}\right]\,\,
\delta \left[(k_0+\mu)^2-m^2-\mathbf{k}^2 \right]
\label{S-prop-time-eB0}
\end{eqnarray}
are the vacuum and matter parts, respectively. After integration of the proper time and making use 
of the identity in Eq.~(\ref{pole-identity}), we find that this is identical to the usual free fermion 
propagator (\ref{free-term}) in the absence of the field. 

Expanding the expression in Eq.~(\ref{S-prop-time-mu1}) to linear order in magnetic field, 
we also easily obtain the following linear in $B$ correction to the fermion propagator:
\begin{eqnarray}
\bar{S}^{(1)}(k) &=& \gamma^1\gamma^2 e B \Bigg\{
\int_{0}^{\infty} s ds\, e^{i s [(k_0+\mu)^2-m^2-\mathbf{k}^2+i\epsilon]} 
+2i\pi  \theta(|\mu|-|k_0|) \theta(-k_0\mu) 
\delta^{\prime}\left[(k_0+\mu)^2-m^2-\mathbf{k}^2 \right]
\Bigg\}\nonumber\\
&&\times \left[(k_0+\mu)\gamma^{0}+m-k^3\gamma^3\right].
\label{S-prop-time-eB1}
\end{eqnarray}
After integration over the proper time and making use of an identity obtained from Eq.~(\ref{pole-identity}) 
by differentiating with respect to ${\cal M}_n^2$, we obtain Eq.~(\ref{linear-term}).

\section{Calculation of the $f_1$, $f_2$, and $f_3$ terms}
\label{App:calculation-j35}

In this appendix, we give the details of the calculation of the radiative corrections to axial 
current due to the $f_1$, $f_2$, and $f_3$ terms. We start from the general form of the 
result in Eq.~(\ref{chiral-current-new}) and calculate separately the two contributions, 
$f_1$ and $f_2+f_3$. At the end we combine all contributions and calculate the final 
result for the $f_1+f_2+f_3$ contribution.

\subsection{Calculation of $f_1$}

Starting from the definition in Eq.~(\ref{Mat-f-1}), we find it convenient to rewrite the expression for $f_1$
in the following equivalent form:
\begin{equation}
f_1 \equiv f_1(m_\gamma)-f_1(\Lambda),
\label{f_1-reg}
\end{equation}
where we took into account that the photon propagator is defined by Eq.~(\ref{photon-propagator}),
with $\Lambda$ playing the role of the ultraviolet regulator. As follows from the definition, 
\begin{eqnarray}
f_1(m_\gamma) &=& - 64 i \pi^2 \alpha eB \frac{\partial }{\partial (m_c)^2} 
\int\frac{d^4 p d^4 k}{(2\pi)^8}  \frac{ \mu (p_0+\mu) -\mathbf{p}\cdot\mathbf{k}-2m^2 }{[(P-K)^2-m_\gamma^2](P^2-m^2)}
\delta \left[\mu^2-m_c^2-\mathbf{k}^2\right]\delta(k_0)\nonumber\\
&=& \frac{16 i \pi \alpha eB}{k_{F}} \frac{\partial }{\partial k_{F}} \Bigg[k_{F}\int\frac{p^2d p d p_0 d\xi}{(2\pi)^5}
 \frac{ \mu (p_0+\mu) -p k_{F}\xi-2m^2 }{(p_0^2-p^2-k_{F}^2+2pk_{F}\xi-m_\gamma^2)[(p_0+\mu)^2-p^2-m^2]}
\Bigg] ,
\end{eqnarray}
where we integrated over the energy $k_0$, the absolute value of the spacial momentum $k$, and 
all angular coordinates except for the angle $\theta_{kp}$ between $\mathbf{k}$ and $\mathbf{p}$.
We also introduced the following short-hand notations: $k_{F}=\sqrt{\mu^2-m_c^2}$ and 
$\xi=\cos\theta_{kp}$. Note that the auxiliary quantities $m_c$ and $k_{F}$ should be replaced by 
the physical fermion mass $m$ and the Fermi momentum $p_{F}=\sqrt{\mu^2-m^2}$, respectively, 
at the end of the calculation. 

The integral over the energy $p_0$ can be calculated, using the following general result for the 
energy integration: 
\begin{equation}
i\int \frac{\left[X(p_0+\mu)\mu+Y\right]dp_0}{\left(p_0^2-b^2\right)\left[(p_0+\mu)^2-a^2\right]}
=  \frac{\pi}{b}  \Bigg[
\frac{\theta(\mu-a)\left[X(b+\mu)\mu+Y\right]}{\left[(b+\mu)^2-a^2 \right]}
-\frac{\theta(a-\mu)\left[X a \mu^2+(a+b)Y\right]}{a\left[(a+b)^2-\mu^2 \right]}
\Bigg],
\label{App:table-integral}
\end{equation}
where $a=\sqrt{p^2+m^2}$ and $b=\sqrt{p^2+k_F^2-2pk_F\xi+m_\gamma^2}$.
Then we obtain
\begin{eqnarray}
f_1 (m_\gamma) &=& \frac{\alpha eB}{2 \pi^3 k_{F}} \frac{\partial }{\partial k_{F}} 
\int\frac{k_{F} p^2d p d\xi}{b}\Bigg(
\frac{\theta(p_F-p)\left[\mu(b+\mu)-p k_{F}\xi-2m^2\right]}{(b+\mu)^2-a^2}
\nonumber\\
&-&
\frac{\theta(p-p_F)\left[a\mu^2-(a+b)(p k_{F}\xi+2m^2)\right]}{a\left[(a+b)^2-\mu^2\right]}
\Bigg)
\label{f1-1st} ,
\end{eqnarray}
The integral over the angular coordinate $\xi$ can be easily performed, leading to the 
following result:
\begin{eqnarray}
f_1(0)  &=& \frac{\alpha eB}{4 \pi^3 k_{F}} \frac{\partial }{\partial k_{F}} 
\int p d p \Bigg[\theta(p_F-p)\Bigg(p+k_F-|p-k_F|+\frac{\mu^2-3m^2-k_F^2}{2a}
\ln\frac{(\mu+|p-k_F|+a)(\mu+p+k_F-a)}{(\mu+p+k_F+a)(\mu+|p-k_F|-a)}
\Bigg)
\nonumber\\
&&+ \theta(p-p_F)\Bigg(p+k_F-|p-k_F|-\frac{2k_Fp}{a}
+\frac{\mu^2-3m^2-k_F^2}{2a}\ln\frac{(a+|p-k_F|)^2 -\mu^2}{(a+p+k_F)^2 -\mu^2 }
\Bigg)\Bigg].
\label{f1-2nd} 
\end{eqnarray}
Here, without loss of generality, we presented the result only for the case of the vanishing 
photon mass. This is justified because, as we will see below, the limit $m_\gamma\to 0$ does 
not produce any infrared singularities in the final result for $f_1$. If needed, an analogous 
expression for the case of a nonzero photon mass $m_\gamma$ can be readily written down
as well. It can be obtained from the above result by making the following three replacements: 
(i) $|p-k_F|\to \sqrt{(p-k_F)^2+m_\gamma^2}$, 
(ii) $p+k_F\to \sqrt{(p+k_F)^2+m_\gamma^2}$, and 
(iii) $\mu^2-3m^2-k_F^2\to \mu^2-3m^2-k_F^2-m_\gamma^2$ at two places in front 
of the logarithms.

After calculating the derivative with respect to $k_{F}$ in Eq.~(\ref{f1-2nd}) and then substituting 
$k_F\to p_F$, we obtain
\begin{eqnarray}
f_1(0)  &=& \frac{\alpha eB}{4 \pi^3 p_{F}} 
\int p d p \Bigg[\theta(p_{F}-p)\Bigg(
\frac{2m^2p}{(p_{F}+\mu)(p_{F}^2-p^2)}
-\frac{p_{F}}{a}\ln\frac{(\mu+p_{F})^2-(p-a)^2}{(\mu+p_{F})^2-(p+a)^2}
\Bigg)
\nonumber\\
&&+ \theta(p-p_{F})\Bigg( 2-\frac{2p}{a}-\frac{p_{F}}{a}\ln\frac{p-p_{F}}{p+p_{F}}
-\frac{2m^2p_{F}^2}{a(a+p)(p^2-p_{F}^2)}
+\frac{2m^2p}{a(p^2-p_{F}^2)}
\Bigg)\Bigg].
\label{f1-3rd} 
\end{eqnarray}
It is easy to check that the above expression has a logarithmic ultraviolet divergency, i.e.,
\begin{equation}
f_1^{\rm UV}(0)  \simeq \alpha eB\frac{2\mu^2+m^2}{4 \pi^3 \sqrt{\mu^2-m^2}} \int \frac{d p}{p}.
\label{f_1_UV-divergency}
\end{equation}
This confirms that an ultraviolet regularization is required in the calculation. As mentioned earlier, 
we utilize the Feynman regularization (\ref{f_1-reg}), which is equivalent to using the photon 
propagator in Eq.~(\ref{photon-propagator}). This is the same regularization, which is commonly 
used in the calculation of vacuum diagrams in QED, when the regularized expression is obtained 
from the divergent one by subtracting the contribution with a large photon mass $\Lambda$. 
In the case at hand, therefore, we need the explicit expression for the function $f_1(\Lambda)$. 
The corresponding calculation is tedious, but straightforward. The result reads
\begin{eqnarray}
f_1 (\Lambda) &=& \frac{\alpha eB}{4\pi^3 p_{F}} 
\int p d p \Bigg[\theta(p_{F}-p)\Bigg(
\frac{p+p_{F}}{\sqrt{(p+p_{F})^2+\Lambda^2}}
+\frac{p-p_{F}}{\sqrt{(p-p_{F})^2+\Lambda^2}}
\nonumber\\
&&+\frac{(p_{F}-p)(2m^2+\Lambda^2)}{\sqrt{(p-p_{F})^2+\Lambda^2} 
\left(2p_{F}^2-2p_{F} p+\Lambda^2+2\mu \sqrt{(p-p_{F})^2+\Lambda^2}\right)}
\nonumber\\
&&-\frac{(p_{F}+p)(2m^2+\Lambda^2)}{\sqrt{(p+p_{F})^2+\Lambda^2} 
\left(2p_{F}^2+2p_{F} p+\Lambda^2+2\mu \sqrt{(p+p_{F})^2+\Lambda^2}\right)}
\nonumber\\
&&+\frac{p_{F}}{a}\ln\frac
{\left(\mu+\sqrt{(p+p_{F})^2+\Lambda^2} +a\right)\left(\mu+\sqrt{(p-p_{F})^2+\Lambda^2} -a\right)}
{\left(\mu+\sqrt{(p-p_{F})^2+\Lambda^2} +a\right)\left(\mu+\sqrt{(p+p_{F})^2+\Lambda^2} -a\right)}
\Bigg)
\nonumber\\
&&+ \theta(p-p_{F})\Bigg( \frac{p+p_{F}}{\sqrt{(p+p_{F})^2+\Lambda^2}}
+\frac{p-p_{F}}{\sqrt{(p-p_{F})^2+\Lambda^2}}-\frac{2p}{a}
\nonumber\\
&&+\frac{(2m^2+\Lambda^2)(p-p_{F})\left(a+\sqrt{(p-p_{F})^2+\Lambda^2}\right)}
{a\sqrt{(p-p_{F})^2+\Lambda^2}\left(2p^2-2pp_{F}+\Lambda^2+2a\sqrt{(p-p_{F})^2+\Lambda^2}\right)}
\nonumber\\
&&+\frac{(2m^2+\Lambda^2)(p+p_{F})\left(a+\sqrt{(p+p_{F})^2+\Lambda^2}\right)}
{a\sqrt{(p+p_{F})^2+\Lambda^2}\left(2p^2+2pp_{F}+\Lambda^2+2a\sqrt{(p+p_{F})^2+\Lambda^2}\right)}
\nonumber\\
&&+\frac{p_{F}}{a}\ln
\frac{2p^2+2pp_{F}+\Lambda^2+2a\sqrt{(p+p_{F})^2+\Lambda^2}}
{2p^2-2pp_{F}+\Lambda^2+2a\sqrt{(p-p_{F})^2+\Lambda^2}} 
\Bigg)
\Bigg].
\label{f1-3rd-gamma} 
\end{eqnarray}
Finally, as follows from the definition in Eq.~(\ref{f_1-reg}),  the regularized expression of $f_1$ reads
\begin{eqnarray}
f_1 &=&  \frac{\alpha eB}{4\pi^3 k_{F}} 
\int_{0}^{p_{F}} p d p \Bigg(
\frac{m^2}{(p_{F}-p)\left(p_{F}+\mu\right)}
-\frac{ m^2 }{(p+p_{F})\left(p_{F}+\mu \right)}
+\frac{p_{F}}{a}\ln\frac
{\left(\mu+p_{F}+p +a\right)\left(\mu+p_{F}-p -a\right)}
{\left(\mu+p_{F}-p +a\right)\left(\mu+p_{F}+p -a\right)}
\Bigg)
\nonumber\\
&&+ \frac{\alpha eB}{4\pi^3 k_{F}} 
\int_{p_{F}}^{\infty} p d p  \Bigg( 2
-\frac{p+p_{F}}{\sqrt{(p+p_{F})^2+\Lambda^2}}
-\frac{p-p_{F}}{\sqrt{(p-p_{F})^2+\Lambda^2}} 
\nonumber\\
&&
+\frac{m^2 \left(a+p-p_{F}\right)}{a(p-p_{F})\left(a+p\right)}
-\frac{(2m^2+\Lambda^2)(p-p_{F})\left(a+\sqrt{(p-p_{F})^2+\Lambda^2}\right)}
{a\sqrt{(p-p_{F})^2+\Lambda^2}\left(2p^2-2pp_{F}+\Lambda^2+2a\sqrt{(p-p_{F})^2+\Lambda^2}\right)}
\nonumber\\
&&
+\frac{m^2\left(a+p+p_{F}\right)} {a(p+p_{F})\left(a+p\right)}
-\frac{(2m^2+\Lambda^2)(p+p_{F})\left(a+\sqrt{(p+p_{F})^2+\Lambda^2}\right)}
{a\sqrt{(p+p_{F})^2+\Lambda^2}\left(2p^2+2pp_{F}+\Lambda^2+2a\sqrt{(p+p_{F})^2+\Lambda^2}\right)}
\nonumber\\
&&+\frac{p_{F}}{a}\ln\frac{p+p_{F}}{p-p_{F}} 
-\frac{p_{F}}{a}\ln
\frac{2p^2+2pp_{F}+\Lambda^2+2a\sqrt{(p+p_{F})^2+\Lambda^2}}
{2p^2-2pp_{F}+\Lambda^2+2a\sqrt{(p-p_{F})^2+\Lambda^2}} 
\Bigg)\Bigg|_{\Lambda\gg\mu}.
\label{f1-3rd-Lambda} 
\end{eqnarray}
Note that, in the first integral below the Fermi surface ($p\leq p_F$), we took the limit 
$\Lambda\to \infty$ because it does not cause any problem. It is essential, however, to keep $\Lambda$ 
finite in the second integral above the Fermi surface ($p\geq p_F$).

A careful analysis of the regularized expression for $f_1$ in Eq.~(\ref{f1-3rd-Lambda}) reveals
a potentially serious problem: both integrals below and above the Fermi surface have infrared 
logarithmic divergencies coming from the regions near $p_F$. These divergencies cannot 
be avoided even when the photon mass is introduced as a regulator. (The divergencies do 
happen to vanish in the theory with massless fermions, $m=0$, but this is of no importance 
as we discuss below.) Fortunately, as we will see below, the corresponding divergencies exactly 
cancel similar infrared divergencies in the expression for $f_2+f_3$. Therefore, we come to the 
conclusion that the appearance of infrared divergencies in $f_1$, as well as in 
$f_2+f_3$, is purely accidental and has no implications on physical observables. They can be 
viewed as a consequence of an ambiguous split of the finite expression $f_1+f_2+f_3$ into two 
separate contributions. 

In order to carefully sort out the cancellation of the above-mentioned (unphysical) infrared 
divergencies, it is useful to explicitly separate the divergent terms from regular ones in the 
corresponding expression for $f_1=f_1^{\rm (IR,div)}+f_1^{\rm (IR,reg)}$. The divergent part of the 
expression reads 
\begin{eqnarray}
f_1^{\rm (IR,div)}&=&\frac{\alpha eBm^2}{4\pi^3 k_{F}}\left[
\int_{0}^{k_F-\epsilon_1} \frac{p d p}{(k_F-p)(k_F+\mu)}
+ \int_{k_F+\epsilon_2}^{\infty} p d p\left(\frac{1}{(p-k_F)(a+p)}-\frac{1}{2p^2}\right)\right]
\nonumber\\
&= & \frac{\alpha eBm^2}{4\pi^3}\left[
\frac{1}{k_F+\mu}\left(\ln\frac{k_F}{\epsilon_1}-1\right)
+\frac{1}{k_F+\mu} \left(\ln\frac{\mu}{\epsilon_2}-\frac{3}{2}\right)
+\frac{1}{2k_F}  \left(\ln\frac{2k_F}{k_F+\mu}+\frac{1}{2}\right)
+\frac{\mu+k_F}{m^2}  \ln\frac{2\mu}{k_F+\mu} \right],
\end{eqnarray}
where we introduced infrared regulators $\epsilon_1$ and $\epsilon_2$ (with $\epsilon_{1},\epsilon_{2} \to 0$)
that allow us to deal with the problem in a rigorous way. Notice that, in the second integral we added a simple 
regular term, whose only purpose is to ensure the ultraviolet convergence of the whole 
expression. The remaining regular part of the expression for $f_1$ reads
\begin{eqnarray}
f_1^{\rm (IR,reg)}&=&\frac{\alpha eB}{4\pi^3 k_{F}} 
\int_{0}^{k_F} p d p \Bigg(
-\frac{ m^2 }{(p+k_F)\left(k_F+\mu \right)}
+\frac{k_F}{a}\ln\frac
{\left(\mu+k_F+p +a\right)\left(\mu+k_F-p -a\right)}
{\left(\mu+k_F-p +a\right)\left(\mu+k_F+p -a\right)}
\Bigg)
\nonumber\\
&&+ \frac{\alpha eB}{4\pi^3 k_{F}} 
\int_{k_F}^{\infty} p d p  \Bigg( 2
-\frac{p+k_F}{\sqrt{(p+k_F)^2+\Lambda^2}}
-\frac{p-k_F}{\sqrt{(p-k_F)^2+\Lambda^2}} 
+\frac{m^2 }{2p^2}
+\frac{2m^2}{a\left(a+p\right)}
+\frac{m^2 } {(p+k_F)\left(a+p\right)}
\nonumber\\
&&
-\frac{(2m^2+\Lambda^2)(p-k_F)\left(a+\sqrt{(p-k_F)^2+\Lambda^2}\right)}
{a\sqrt{(p-k_F)^2+\Lambda^2}\left(2p^2-2pk_F+\Lambda^2+2a\sqrt{(p-k_F)^2+\Lambda^2}\right)}
\nonumber\\
&&
-\frac{(2m^2+\Lambda^2)(p+k_F)\left(a+\sqrt{(p+k_F)^2+\Lambda^2}\right)}
{a\sqrt{(p+k_F)^2+\Lambda^2}\left(2p^2+2pk_F+\Lambda^2+2a\sqrt{(p+k_F)^2+\Lambda^2}\right)}
\nonumber\\
&&+\frac{k_F}{a}\ln\frac{p+k_F}{p-k_F} 
-\frac{k_F}{a}\ln
\frac{2p^2+2pk_F+\Lambda^2+2a\sqrt{(p+k_F)^2+\Lambda^2}}
{2p^2-2pk_F+\Lambda^2+2a\sqrt{(p-k_F)^2+\Lambda^2}} 
\Bigg)\Bigg|_{\Lambda\gg\mu} .
\end{eqnarray}
Calculating the integrals in the case $m\ll |\mu|$, we arrive at the following 
results:
\begin{eqnarray}
f_1^{\rm (IR,div)} &\simeq & \frac{\alpha eBm^2}{8\pi^3\mu}\left(
\ln\frac{\mu}{\epsilon_1} 
+ \ln\frac{\mu}{\epsilon_2}  -1\right) , \\
f_1^{\rm (IR,reg)}&\simeq & \frac{\alpha eB \mu}{2\pi^3}\left(\ln\frac{\Lambda}{2\mu}+\frac{5}{4}\right)
+\frac{\alpha eBm^2}{2\pi^3\mu}\ln\frac{\Lambda}{2\mu}.
\end{eqnarray}

\subsection{Calculation of $f_2+f_3$}

In this subsection, we calculate the expression for the sum $f_2+f_3$ starting from the definition 
in Eqs.~(\ref{Mat-f-2}) and (\ref{Mat-f-3}). As we we see below, the corresponding expression has
no ultraviolet divergencies. Therefore, we could take the limit $\Lambda \to \infty$ in the 
expression for $f_2+f_3$, i.e.,
\begin{eqnarray}
f_2 +f_3 &=&   \frac{64 i \pi^2 \alpha eB}{3} \frac{\partial }{\partial (m_a)^2} 
\int\frac{d^4 p d^4 k}{(2\pi)^8}  \frac{
3(p_0+\mu)^2-3\mu(p_0+\mu) +p^2-pk_F \xi +3m^2}{\left[(P-K)^2-m_\gamma^2\right](P^2-m_a^2)}
\delta \left(\mu^2-m^2-\mathbf{k}^2\right)\delta(k_0)
\nonumber\\
&=&\frac{ 32 i \pi \alpha eB k_F }{3} \frac{\partial }{\partial (m_a)^2} \int\frac{p^2d p d p_0 d\xi}{(2\pi)^5}
\frac{4p^2-pk_F \xi +3m_a^2+3m^2-3\mu(p_0+\mu)}{(p_0^2-p^2-k_F^2+2pk_F\xi-m_\gamma^2)[(p_0+\mu)^2-p^2-m_a^2]} ,
\end{eqnarray}
where we integrated over the energy $k_0$, the absolute value of the spacial momentum $k$, and 
all angular coordinates except for the angle $\theta_{kp}$ between $\mathbf{k}$ and $\mathbf{p}$.
We also introduced an auxiliary quantity $m_a$, which should be replaced by the physical fermion 
mass $m$ at the end of the calculation. It should be also noted that some terms independent of 
$m_a$ were dropped in the integrand of the last expression. This is justified because they vanish 
anyway after the derivative with respect to $m_a^2$ is calculated. 

In order to calculate the integral over the energy $p_0$, we use again the result in Eq.~(\ref{App:table-integral}).
Then, we arrive at 
\begin{eqnarray}
f_2 +f_3 &=& -\frac{\alpha eB}{6 \pi^3} \frac{\partial }{\partial p_F} 
\int\frac{p^2d p d\xi}{b}\Bigg(
\frac{\theta(p_F-p)\left[-3b\mu+4p^2-pk_F \xi -3p_F^2+3m^2\right]}{ (b+\mu)^2-a^2 }
\nonumber\\
&-&\frac{\theta(p-p_F)\left[3b \mu^2+(a+b)(4p^2-pk_F \xi -3p_F^2+3m^2)\right]}{a\left[(a+b)^2-\mu^2 \right]}
\Bigg),
\label{f23-1st}
\end{eqnarray}
where $b=\sqrt{p^2+k_F^2-2pk_F\xi+m_\gamma^2}$ and $a=\sqrt{p^2+\mu^2-p_F^2}$. Note that, in this
subsection, we distinguish the quantity $p_F\equiv \sqrt{\mu^2-m_a^2}$ from the physical Fermi
momentum $k_F=\sqrt{\mu^2-m^2}$. After the partial derivative with respect to $p_F$ is performed, 
the value of $p_F$ will be replaced by $k_F$. Similarly, the auxiliary quantity $a$, which is a function 
of $p_F$, will be replaced by its physical counterpart $\sqrt{p^2+m^2}$.

The integral over the angular coordinate $\xi$ can be easily performed, leading to the following result:
\begin{eqnarray}
f_2 +f_3 &=& -\frac{\alpha eB}{12\pi^3 k_F} \frac{\partial }{\partial p_F} 
\int p d p \Bigg[\theta(p_F-p)\Bigg(p+k_F-|p-k_F|
+4\mu\ln\frac{(\mu+|p-k_F|)^2 -a^2}{(\mu+p+k_F)^2 -a^2 }
\nonumber\\
&&+\frac{7(k_F^2-p_F^2)+8p^2+14m^2}{2a}\ln\frac{(\mu+|p-k_F|+a)(\mu+p+k_F-a)}{(\mu+p+k_F+a)(\mu+|p-k_F|-a)}
\Bigg)
\nonumber\\
&&+ \theta(p-p_F)\Bigg(p+k_F-|p-k_F|-\frac{2k_Fp}{a}
+4\mu\ln\frac{(a+|p-k_F|+\mu)(a+p+k_F-\mu)}{(a+p+k_F+\mu)(a+|p-k_F|-\mu)}
\nonumber\\
&&+\frac{7(k_F^2-p_F^2)+8p^2+14m^2}{2a}\ln\frac{(a+|p-k_F|)^2 -\mu^2}{(a+p+k_F)^2 -\mu^2 }
\Bigg)\Bigg],
\label{f23-2nd}
\end{eqnarray}
where, without loss of generality, we again took the limit of vanishing photon mass. 

It is straightforward to show that the integral in Eq.~(\ref{f23-2nd}) has infrared logarithmic 
divergencies, similar to those in function $f_1$. Therefore, we follow the same kind of analysis 
as in the case of $f_1$ and extract explicitly the following infrared divergent terms: 
\begin{eqnarray}
f_2^{\rm (IR,div)} +f_3^{\rm (IR,div)} &=& -\frac{\alpha eB}{12\pi^3 k_F} \frac{\partial }{\partial p_F} 
\int p d p \Bigg[ \theta(p_F-\epsilon_1-p)\Bigg(-\frac{3m^2}{\mu}\ln\frac{(\mu+|p-k_F|)^2 -a^2}{(\mu+p+k_F)^2 -a^2 }
\Bigg)\nonumber\\
&&\hspace{1.4in}+ \theta(p-p_F-\epsilon_2)\Bigg(\frac{3m^2}{a}\ln\frac{a-\mu+p-k_F}{a+\mu+p+k_F}+\frac{3m^2(k_F+\mu)}{p^2}
\Bigg) \Bigg]\nonumber\\
&=&\frac{\alpha eB m^2}{4 \pi^3} \Bigg[\frac{1}{k_F+\mu} \left(\ln\frac{\mu \epsilon_1}{k_F(k_F+\mu)}-\frac{2k_F}{\mu}\right)
+\frac{1}{k_F+\mu}\ln\frac{\epsilon_2}{k_F+\mu} +\frac{k_F+\mu}{k_F^2} \Bigg].
\end{eqnarray}
Then, the leftover regular part reads
\begin{eqnarray}
f_2^{\rm (IR,reg)} +f_3^{\rm (IR,reg)} &=& -\frac{\alpha eB}{12\pi^3 k_F} \frac{\partial }{\partial p_F} 
\int p d p \Bigg[\theta(p_F-p)\Bigg(p+k_F-|p-k_F|
+4\mu\ln\frac{(\mu+|p-k_F|)^2 -a^2}{(\mu+p+k_F)^2 -a^2 }\nonumber\\
&&+\frac{7(k_F^2-p_F^2)+8p^2+14m^2}{2a}\ln\frac{(\mu+|p-k_F|+a)(\mu+p+k_F-a)}{(\mu+p+k_F+a)(\mu+|p-k_F|-a)}
+\frac{3m^2}{\mu}\ln\frac{(\mu+|p-k_F|)^2 -a^2}{(\mu+p+k_F)^2 -a^2 }
\Bigg)
\nonumber\\
&&+ \theta(p-p_F)\Bigg(p+k_F-|p-k_F|-\frac{2k_Fp}{a}
+4\mu\ln\frac{(a+|p-k_F|+\mu)(a+p+k_F-\mu)}{(a+p+k_F+\mu)(a+|p-k_F|-\mu)}
\nonumber\\
&&+\frac{7(k_F^2-p_F^2)+8p^2+14m^2}{2a}\ln\frac{(a+|p-k_F|)^2 -\mu^2}{(a+p+k_F)^2 -\mu^2 }
-\frac{3m^2}{a}\ln\frac{a-\mu+p-k_F}{a+\mu+p+k_F}-\frac{3m^2(k_F+\mu)}{p^2}
\Bigg)\Bigg].
\end{eqnarray}
After calculating the derivative with respect to $p_F$ in the regular piece (note that $a=\sqrt{p^2+\mu^2-p_F^2}$, 
i.e., $a$ is a function of $p_F$) and substituting $p_F\to k_F$ afterwards, we obtain
\begin{eqnarray}
f_2^{\rm (IR,reg)} +f_3^{\rm (IR,reg)} &=&-\frac{\alpha eB}{2 \pi^3} \left(
\frac{k_F^2}{3\mu}
+\frac{m^2}{\mu}\ln\frac{k_F^2}{ \mu(\mu+k_F)} 
+\frac{m^2(k_F+\mu)}{2k_F^2}
\right)\nonumber\\
&+&\frac{\alpha eB}{4\pi^3} 
\int \frac{p d p}{a^2} \Bigg[
\theta(k_F-p)  \Bigg( \frac{2m^2 p}{\mu(k_F+\mu)}
+\frac{p^2}{a}\ln\frac{k_F(k_F+\mu)+p(a-p)}{k_F(k_F+\mu)-p(a+p)}\Bigg)
\nonumber\\
&&+ \theta(p-k_F)    \Bigg(\frac{2pk_F}{3a}+\frac{8a k_F}{3(a+p)}-\frac{m^2(a-p)}{a k_F+p\mu}
+\frac{p^2}{a}\ln\frac{p-k_F}{p+k_F}+\frac{m^2}{a}\ln\frac{a+p-k_F-\mu}{a+p+k_F+\mu}
\Bigg)\Bigg] .
\end{eqnarray}
As is easy to check, the integrand is $\propto 1/p^3$ when $p\to \infty$. It is clear, therefore, that 
the expression is convergent and no additional ultraviolet regularization is needed. 
Integrating over the momentum, we finally obtain
\begin{eqnarray}
f_2^{\rm (IR,reg)} +f_3^{\rm (IR,reg)} &=&-\frac{\alpha eBk_F}{2 \pi^3} \Bigg[ 
\frac{k_F}{3\mu}
+\frac{m^2}{k_F \mu}\ln\frac{k_F^2}{ \mu(\mu+k_F)} 
+\frac{m^2(k_F+\mu)}{2k_F^3}
\Bigg]\nonumber\\
&+&\frac{\alpha eB}{2 \pi^3} \Bigg[ k_F\Bigg( 1+\frac{m^2}{\mu(k_F+\mu)} -\frac{\mu^2+m^2}{k_F \mu}\ln\frac{k_F+\mu}{\mu}\Bigg)
\nonumber\\
&+&\mu\Bigg(\frac{ \mu-k_F }{k_F+\mu}\left(1+\frac{k_F m^2}{3\mu^2(k_F+\mu)}\right)
+\frac{k_F(k_F+\mu)}{2\mu^2}\ln\frac{\mu}{k_F}-\left(2+\frac{k_F}{\mu}\right)\ln\frac{2\mu}{k_F+\mu}
+\ln 2-1
 \Bigg) \Bigg].
\end{eqnarray}
For $m\ll|\mu|$, the final expressions for the infrared divergent and regular 
contributions simplify as follows:  
\begin{eqnarray}
f_2^{\rm (IR,div)} +f_2^{\rm (IR,div)}  &\simeq &  
\frac{\alpha eBm^2}{2 \pi^3 \mu} \Bigg[\frac{1}{4} \left(\ln\frac{\epsilon_1}{2\mu}- 2\right)
+\frac{1}{4}\ln\frac{\epsilon_2}{2\mu} + 1 \Bigg],  \\
f_2^{\rm (IR,reg)} +f_3^{\rm (IR,reg)} &\simeq &  -\frac{\alpha eB\mu}{6 \pi^3} 
- \frac{\alpha eBm^2}{24\pi^3 \mu} .
\end{eqnarray}

\subsection{Collecting all contributions}

As seen from Eq.~(\ref{chiral-current-new}), the final expression for the axial current is given in terms of 
the sum $f_1+f_2+f_3$. The corresponding function is obtained by collecting all the divergent and regular 
terms calculated in the previous two subsections of this appendix. In the case $m\ll|\mu|$, in 
particular, the result reads
\begin{eqnarray}
f_1+f_2+f_3 &=&  
f_1^{\rm (IR,div)} + f_2^{\rm (IR,div)} +f_2^{\rm (IR,div)}  
+f_1^{\rm (IR,reg)}+f_2^{\rm (IR,reg)} +f_3^{\rm (IR,reg)} \nonumber\\
&\simeq &  \frac{\alpha eB\mu}{2\pi^3} \left(\ln\frac{\Lambda}{2\mu}+\frac{11}{12}\right)
+\frac{\alpha eBm^2}{2 \pi^3 \mu} \left(\ln\frac{\Lambda}{2^{3/2}\mu}+\frac{1}{6}\right).
\end{eqnarray}
Notice that all infrared regulators ($\epsilon_1$ and $\epsilon_2$), which were introduced 
in the divergent parts of $f_1$ and $f_2+f_3$ canceled out. The only regulator in the
last expression is the ultraviolet one $\Lambda$. In the final expression for the axial 
current (\ref{j53-small-m-final}), this dependence on the ultraviolet regulator cancels 
out exactly with a similar dependence coming from the counterterms contribution in 
Eq.~(\ref{counterterm-contribution-general}).


\begin{thebibliography}{99}


\bibitem{Kharzeev:2004ey} 
  D.~Kharzeev,
  Phys.\ Lett.\ {\bf B633}, 260 (2006).

\bibitem{Kharzeev:2007tn} 
  D.~Kharzeev and A.~Zhitnitsky,
  Nucl.\ Phys.\ {\bf A797}, 67 (2007).

  \bibitem{Kharzeev:2007jp} 
  D.~E.~Kharzeev, L.~D.~McLerran, and H.~J.~Warringa,
  Nucl.\ Phys.\ {\bf A803}, 227 (2008).

\bibitem{Fukushima:2008xe} 
  K.~Fukushima, D.~E.~Kharzeev, and H.~J.~Warringa,
  Phys.\ Rev.\ D {\bf 78}, 074033 (2008).
  
  \bibitem{Fukushima:2012vr} 
  K.~Fukushima, Lect. Notes Phys. {\bf 871}, 241 (2013).

\bibitem{collisions} 
B.~I.~Abelev {\it et al.}  (STAR Collaboration),
  Phys.\ Rev.\ Lett.\  {\bf 103}, 251601 (2009);
  Phys.\ Rev.\ C {\bf 81}, 054908 (2010).

\bibitem{Adamczyk:2013kcb} 
  L.~Adamczyk {\it et al.}  (STAR Collaboration),
  arXiv:1303.0901.

\bibitem{Wang:2012qs} 
  G.~Wang (STAR Collaboration), Nucl. Phys. {\bf A904-905}, 248c (2013).
  
\bibitem{Ke:2012qb} 
    H.~Ke (STAR Collaboration),
  J.\ Phys.\ Conf.\ Ser.\  {\bf 389}, 012035 (2012).
  
  \bibitem{Selyuzhenkov:2011xq} 
  I.~Selyuzhenkov (ALICE Collaboration),
  Prog.\ Theor.\ Phys.\ Suppl.\  {\bf 193}, 153 (2012).
  
  \bibitem{Voloshin:2004vk} 
  S.~A.~Voloshin,
  Phys.\ Rev.\ C {\bf 70}, 057901 (2004).

\bibitem{Kharzeev:2009fn} 
D.~E.~Kharzeev,
  Ann. Phys. (N.Y.)  {\bf 325}, 205 (2010);
K.~Fukushima, D.~E.~Kharzeev and H.~J.~Warringa,
  Nucl.\ Phys.\ {\bf A836}, 311 (2010).

\bibitem{Vilenkin:1980ft} 
  A.~Vilenkin,
  Phys.\ Rev.\ D {\bf 22}, 3080 (1980).

\bibitem{Zhitnitsky}
  M.~A.~Metlitski and A.~R.~Zhitnitsky,
  Phys.\ Rev.\ D {\bf 72}, 045011 (2005).

\bibitem{Newman}
  G.~M.~Newman and D.~T.~Son,
  Phys.\ Rev.\ D {\bf 73}, 045006 (2006).
  
\bibitem{chiral-shift-3}
  E.~V.~Gorbar, V.~A.~Miransky, and I.~A.~Shovkovy,
  Phys.\ Rev.\ D {\bf 83}, 085003 (2011).

\bibitem{Burnier:2011bf} 
  Y.~Burnier, D.~E.~Kharzeev, J.~Liao, and H.~-U.~Yee,
  Phys.\ Rev.\ Lett.\  {\bf 107}, 052303 (2011);
  arXiv:1208.2537.
    
\bibitem{ABJ} 
S.~L.~Adler,
  Phys.\ Rev.\  {\bf 177}, 2426 (1969);
J.~S.~Bell and R.~Jackiw,
  Nuovo Cimento A {\bf 60}, 47 (1969).

\bibitem{Adler:1969er} 
S.~L.~Adler and W.~A.~Bardeen,
  Phys.\ Rev.\  {\bf 182}, 1517 (1969).
  
\bibitem{Basar-Dunne} 
  G.~Basar and G.~V.~Dunne, Lect. Notes Phys. {\bf 871}, 261 (2013).
 
\bibitem{chiral-shift-1} 
  E.~V.~Gorbar, V.~A.~Miransky, and I.~A.~Shovkovy,
  Phys.\ Rev.\ C {\bf 80}, 032801(R) (2009).
  
\bibitem{chiral-shift-2} 
  E.~V.~Gorbar, V.~A.~Miransky, and I.~A.~Shovkovy,
  Phys.\ Lett.\ {\bf B695}, 354 (2011).

\bibitem{Fukushima:2010zza} 
  K.~Fukushima and M.~Ruggieri,
  Phys.\ Rev.\ D {\bf 82}, 054001 (2010).
  
\bibitem{Peskin} M.~E.~Peskin and D.~V.~Schroeder,
{\em An Introduction To Quantum Field Theory},
(Westview Press, Boulder, 1995) 842 pages.  

  \bibitem{Schwinger:1951nm}
  J.~S.~Schwinger,
  Phys.\ Rev.\  {\bf 82}, 664 (1951).

\bibitem{Stueckelberg:1957zz} 
  E.~C.~G.~Stueckelberg,
  Helv.\ Phys.\ Acta {\bf 30}, 209 (1957).
 

\bibitem{Gorbar:2013uga} 
 E.~V.~Gorbar, V.~A.~Miransky, I.~A.~Shovkovy, and X.~Wang,
  arXiv:1306.3245.

\bibitem{footnote-neutrality} As follows from Gauss's law, the consistent description of the 
dynamics in QED at finite density requires the overall neutrality of the system. The neutralizing 
background can be provided, for example, by protons (in neutron stars) or nuclei (in white dwarfs).
This implies that all tadpole diagrams should cancel. In the diagrammatic form of the expression 
for the axial current in Fig.~\ref{fig-correlator}, this corresponds to removing the one-particle 
reducible diagram (not shown in Fig.~\ref{fig-correlator}) that is made of two fermion loops 
connected by a photon line.
 
\bibitem{Gao:2012ix} 
  J.-H.~Gao, Z.-T.~Liang, S.~Pu, Q.~Wang, and X.-N.~Wang,
  Phys.\ Rev.\ Lett.\  {\bf 109}, 232301 (2012);
  J.-W.~Chen, S.~Pu, Q.~Wang, and X.-N.~Wang,
  arXiv:1210.8312.

\bibitem{Weinberg} S.~Weinberg, {\em The Quantum Theory of Fields},
(Cambridge University Press, Cambridge, England, 1995) Vol. 1, pp. 564-596. 

\bibitem{Kinoshita:1962ur} 
  T.~Kinoshita,
  J.\ Math.\ Phys.\  {\bf 3}, 650 (1962).

\bibitem{Lee:1964is} 
  T.~D.~Lee and M.~Nauenberg,
  Phys.\ Rev.\  {\bf 133}, B1549 (1964).

\bibitem{Fomin:1976am} 
  P.~I.~Fomin, V.~A.~Miransky, and Y. A.~Sitenko,
  Phys.\ Lett.\ {\bf B64}, 444 (1976).

\bibitem{footnote} 
In massless QED without external electromagnetic fields, fermions interact through neutral vector 
currents \cite{Fomin:1976am}, despite the vanishing renormalized electric charge. In fact, massless QED 
yields the simplest example of an unparticle gauge field theory \cite{Georgi:2007ek}, in which the infrared 
fixed point is Gaussian. There are no one-particle asymptotic states in unparticle theories. 
Instead, the asymptotic states are described by fermionic and bosonic jets 
\cite{Lee:1964is,Fomin:1976am,Georgi:2007ek}.

\bibitem{Georgi:2007ek} 
  H.~Georgi,
  Phys.\ Rev.\ Lett.\  {\bf 98}, 221601 (2007).
  
  \bibitem{Rubakov:2010qi} 
  V.~A.~Rubakov,
  arXiv:1005.1888.
  
\end{thebibliography}
\end{document}